\documentclass[reprint,amsmath,amssymb,pre]{revtex4-2}

\usepackage{graphicx}
\usepackage{color}
\usepackage{dcolumn}
\usepackage{bm}
\usepackage{nicefrac}
\usepackage{hyperref}
\usepackage[textsize=tiny]{todonotes}

\renewcommand{\eqref}[1]{Eq.~(\ref{#1})}
\newcommand{\figref}[1]{Fig.~\ref{#1}}
\newcommand{\figsref}[1]{Figs.~\ref{#1}}
\newcommand{\secref}[1]{Sec.~\ref{#1}}
\newcommand{\secsref}[1]{Secs.~\ref{#1}}
\newcommand{\appref}[1]{Appendix~\ref{#1}}
\newcommand{\refcite}[1]{Ref.~\cite{#1}}

\newcommand{\phiT}{\phi_{\text{T}}}

\begin{document}

\title{Programmable phase behavior in fluids with designable interactions}

\author{Fan Chen}
\affiliation{Department of Chemistry, Princeton University, Princeton, NJ 08544, USA}
\author{William M.~Jacobs}
\email{wjacobs@princeton.edu}
\affiliation{Department of Chemistry, Princeton University, Princeton, NJ 08544, USA}

\date{\today}

\begin{abstract}
  We introduce a method for solving the ``inverse'' phase equilibria problem: How should the interactions among a collection of molecular species be designed in order to achieve a target phase diagram?
  Using techniques from convex optimization theory, we show how to solve this problem for phase diagrams containing a large number of components and many coexisting phases with prescribed compositions.
  We apply our approach to commonly used mean-field models of multicomponent fluids and then use molecular simulations to verify that the designed interactions result in the target phase diagrams.
  Our approach enables the rational design of ``programmable'' fluids, such as biopolymer and colloidal mixtures, with complex phase behavior.
\end{abstract}

\maketitle

\section{Introduction}

The observation that proteins and nucleic acids can demix to form ``biomolecular condensates'' within living cells~\cite{hyman2014liquid,shin2017liquid,alberti2019considerations} has sparked intense interest in understanding the phase behavior of complex, multicomponent fluids~\cite{jacobs2017phase,mao2019phase,jacobs2021self,shrinivas2021phase,carugno2022instabilities,graf2022thermodynamic,zwicker2022evolved}.
Although multicomponent phase equilibria is a foundational topic of chemical physics~\cite{gibbs1878equilibrium}, and many important theoretical contributions\cite{griffiths1970critical,sear2003instabilities,jacobs2013predicting} in this area predate the current popularity of biomolecular condensate research, intracellular phase separation provides compelling motivation for the renewed focus on this question.
In particular, intracellular phase separation can establish \textit{coexisting} condensates with \textit{distinct molecular compositions}, which are required for carrying out specific biological functions~\cite{banani2017biomolecular,ditlev2018who}.
Spontaneous intracellular phase separation is widely believed to be governed primarily by thermodynamics~\cite{brangwynne2015polymer,choi2020physical,dignon2020biomolecular,villegas2022molecular}, even though transport dynamics and nonequilibrium processes may affect the phase behavior observed \textit{in vivo}~\cite{berry2018physical,soding2020mechanisms}.
It is therefore important to understand the relationship between biomolecular interactions and the capacity for biological fluids to self-organize into chemically diverse droplets via thermodynamically-driven phase separation.
Despite considerable progress towards dissecting the molecular determinants of phase separation in biological~\cite{boeynaems2019spontaneous,greig2020arginine}, synthetic~\cite{simon2017programming,lu2020multiphase,kaur2021sequence}, and theoretical~\cite{lin2018theories,harmon2018differential} models with a few distinct species, improved theoretical tools are needed for studying phase equilibria in complex mixtures with a large number of molecular components and more than a handful of coexisting phases.

Mean-field models serve as common starting points for describing multicomponent fluid-phase equilibria~\cite{jacobs2017phase,mao2019phase,jacobs2021self,shrinivas2021phase,carugno2022instabilities,graf2022thermodynamic,zwicker2022evolved}.
The simplest mean-field models that can account for phase transitions invoke the \textit{pairwise} approximation for intermolecular interactions, meaning that the net attractive or repulsive interactions between each pair of molecular species can be described by a single interaction parameter, or coupling coefficient, which is independent of the local molecular concentrations.
The Flory--Huggins~\cite{colby2003polymer} and regular-solution models~\cite{porter2021phase} are commonly used examples of such mean-field models.
Landau free energies~\cite{landau2013statistical} and virial expansions of equations of state~\cite{hansen2013theory} also satisfy this approximation when truncated to lowest order.
Although methods for computing phase coexistence and constructing phase diagrams in binary and ternary fluids are well established~\cite{porter2021phase}, predicting phase coexistence in multicomponent fluids---whose phase diagrams exist in high-dimensional spaces---is both conceptually and algorithmically challenging.
The purpose of this article is to introduce a new approach for solving this problem.

Within the context of a pairwise-interaction model, the central challenge is to map an \textit{interaction matrix} of coupling coefficients to an equilibrium phase diagram, and vice versa.
Given an interaction matrix and the total concentrations of all the molecular species in a mixture, we wish to determine whether phase separation will take place, and, if so, the compositions of the coexisting bulk phases at equilibrium.
The mole fractions of the coexisting bulk phases can then be determined from mass conservation.
We refer to these calculations as solving the \textit{forward problem}.
However, carrying out these calculations can be challenging due to the combinatorial complexity associated with multicomponent phase coexistence.
In order to identify coexisting phases in a theoretical model of a multicomponent mixture, it is first necessary to locate all the candidate phases in the high-dimensional concentration space.
This constitutes a search problem whose complexity scales exponentially with the number of molecular components.
As a result, algorithms for solving the forward problem are often limited to mixtures with a small number of components, or they employ additional assumptions to simplify the search problem.

\begin{figure*}
  \includegraphics[width=0.825\textwidth]{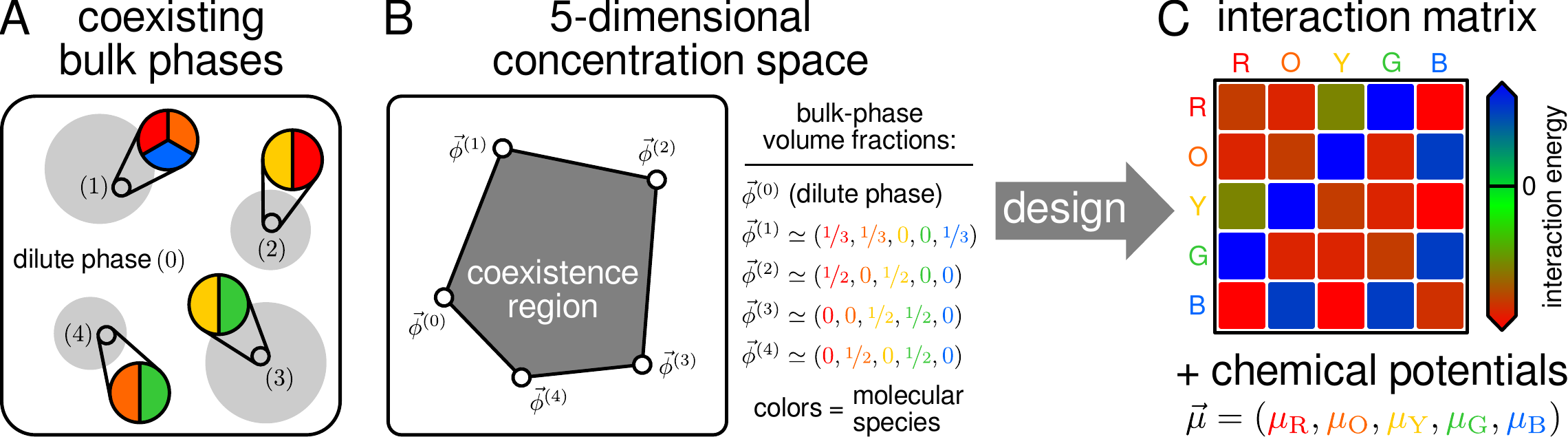}
  \caption{\textbf{Example multicomponent phase-diagram design problem.}
    (A)~Schematic of the design problem.  Each condensed phase (gray circles) has a distinct composition of the five molecular components (represented by colors red, orange, yellow, green, and blue).  The target compositions of these bulk phases are indicated by pie charts.
    (B)~The scenario illustrated in panel A corresponds to a phase diagram in a 5-dimensional concentration space.  The component volume fractions in each of the $K=4$ condensed phases, $\vec\phi^{(\alpha)}$ for $\alpha = 1, \ldots, 4$, specify the design problem.  At equilibrium, any mixture of the five components with a volume-fraction vector inside the convex hull of the coexisting phases (shaded region) will phase-separate into bulk phases with the prescribed molecular compositions.
    (C)~Solving the design problem yields a $5\times5$ interaction matrix, $\bm\epsilon$, and a corresponding chemical-potential vector, $\vec\mu$, that result in coexistence among the $K+1$ bulk phases shown in panel B.}
  \label{fig:1}
\end{figure*}

Here we focus on the \textit{inverse problem}: designing an interaction matrix to achieve a target phase diagram.
Imposing a target phase diagram means that the compositions of the coexisting bulk phases are specified directly as design criteria, while the mole fractions of the coexisting bulk phases can be determined \textit{a priori} for a mixture with specified total concentrations.
Thus, solving the inverse problem provides an alternative approach for associating an equilibrium phase diagram with an interaction matrix, avoiding the need to search for candidate phases.
This approach can also provide insight into the high-dimensional ``design space'' of pairwise interaction matrices that might map to the same, or extremely similar, sets of coexisting phases.
Furthermore, because the time-evolution of complex fluids is often limited by slow kinetics, mapping the relationship between interaction matrices and \textit{metastable} coexisting phases, which may be relevant when a phase-separating fluid reaches a local (but not global) equilibrium, is a similarly important goal.
Inverse-problem approaches are also well suited for designing complex fluids with prescribed metastable phases~\cite{jacobs2021self}.

In this article, we show that an inverse-problem approach can be applied to design equilibrium phase diagrams with arbitrary condensed-phase compositions.
We first explain how this strategy can be applied to generic mean-field models with pairwise intermolecular interactions.
We then show that our approach reveals several unexpected features of multicomponent phase diagrams, demonstrating ways in which high-dimensional phase diagrams can differ qualitatively from the intuitive phase behavior of simple fluids.
Finally, we perform molecular simulations and free-energy calculations to confirm that the predictions of our approach apply beyond mean-field theoretical models.
Our ability to design coexisting phases in non-mean-field simulations suggests that pairwise models, despite their simplicity, can be useful for understanding and manipulating complex phase diagrams of chemically realistic multicomponent fluids.

\section{Results and Discussion}

Throughout this article, we consider incompressible fluids comprising $N$ solute species and a solvent.
We refer to the solute species interchangeably as either the \textit{molecular species} or the \textit{components} in the multicomponent fluid model.
Assuming that the intermolecular interactions in the solution are pairwise additive, the vector of excess chemical potentials for all molecular species can be written in the form
\begin{equation}
  \label{eq:muex}
  \vec\mu_{\text{ex}}(\vec\phi;\bm{\epsilon},\vec v) = \vec\mu_{\text{v}}(\vec\phi;\vec v) + \bm{\epsilon}\vec\phi,
\end{equation}
where $\vec\phi$ and $\vec v$ represent the volume fractions and molecular volumes of the components, respectively, and $\vec\mu_{\text{v}}$ is independent of the symmetric interaction matrix, $\bm{\epsilon}$.
We set the thermal energy $k_{\text{B}}T = 1$ for brevity.
Importantly, we assume that the elements of $\bm{\epsilon}$ are independently tunable throughout this work.
The osmotic pressure, $P(\vec\phi; \bm\epsilon, \vec{v})$, which can be determined from \eqref{eq:muex} via the Gibbs--Duhem relation, is also linear with respect to $\bm{\epsilon}$.
\eqref{eq:muex} describes the mean-field Flory--Huggins~\cite{colby2003polymer}, regular solution~\cite{porter2021phase}, and van der Waals~\cite{hansen2013theory} models, as well as approximate field-theoretic treatments of sequence-dependent heteropolymer mixtures~\cite{wessen2022analytical}, with appropriate choices of $\vec\mu_{\text{v}}$.
This formalism takes the chemical potential of the solvent and all pairwise interactions involving the solvent to be zero; we make this choice without loss of generality, since the reference states for the solvent and all pure components do not affect the phase equilibria of incompressible mixtures~\cite{colby2003polymer}.
We further note that the class of mean-field models represented by \eqref{eq:muex} only describes the average molecular concentrations within each phase, as $\vec\phi$ is the sole independent variable.

\subsection{Inverse design of phase equilibria in mixtures with pairwise interactions}
\label{sec:inverse-design}

\begin{figure*}
  \includegraphics[width=\textwidth]{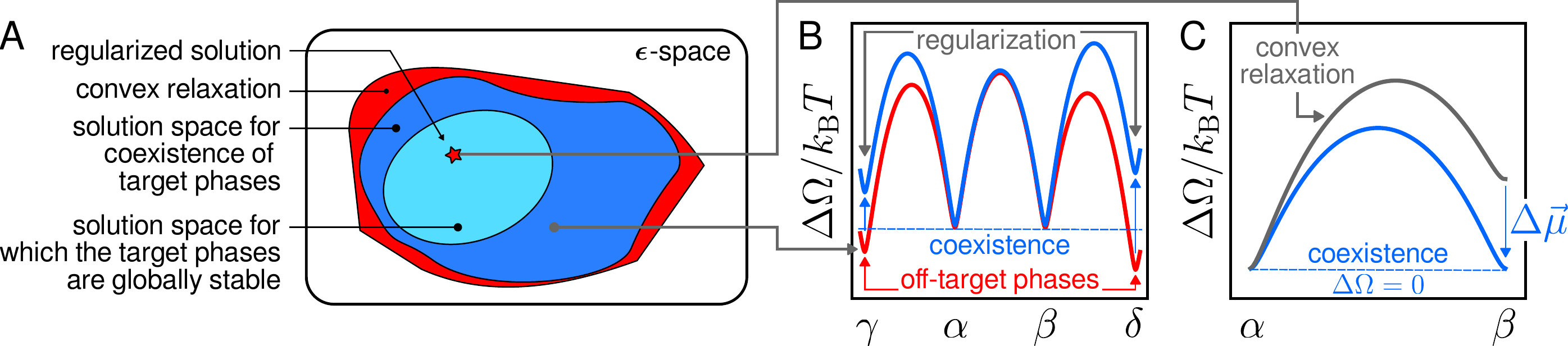}
  \caption{\textbf{Inverse-design approach to multicomponent phase coexistence.}
    (A)~Schematic of the $N(N+1)/2$-dimensional interaction-matrix ``$\bm\epsilon$-space''.  Given a feasible design problem (e.g., \figref{fig:1}B), there exists a ``solution space'' (dark blue region) containing all interaction matrices that result in coexistence among the bulk phases specified by the target phase diagram.  If the design problem is infeasible, then this solution space does not exist.  Every interaction matrix within the solution space has a corresponding chemical potential vector at phase coexistence.  Solutions to the convex relaxation (red region) may lie outside the solution space.  Within the solution space, there may be a subspace for which the target phases are globally stable (cyan region; see text for details).
    (B)~Schematic of the regularization heuristic for eliminating globally stable (i.e., $\Delta\Omega \le 0$) off-target phases, illustrated here for target phases $\alpha$ and $\beta$ and representative off-target phases $\gamma$ and $\delta$.  Regularization is used to bias solutions toward the globally-stable subspace depicted in panel A (see text for details).
    (C)~Due to the approximations employed in constructing the convex relaxation, the chemical potential vector that is obtained via convex optimization may not yield precise phase coexistence (i.e., zero difference in the grand potential, $\Delta\Omega$, between phases).  We therefore perform a common tangent plane construction to correct the coexistence chemical potentials that are found via convex optimization.  This procedure is shown schematically for a pair of target phases $\alpha$ and $\beta$.  Two grand-potential surfaces, projected along a path through concentration space between $\vec\phi^{(\alpha)}$ and $\vec\phi^{(\beta)}\!$, are depicted before (gray curve) and after (blue curve) computing the correction to the chemical-potential vector, $\Delta\vec\mu$.}
  \label{fig:2}
\end{figure*}

Our objective is to find an $N \times N$ interaction matrix, $\bm{\epsilon}$, and an $N$-dimensional chemical potential vector, $\vec\mu$, that lead to phase coexistence among a dilute phase and $K$ condensed phases.
An inverse-design problem is defined by the target volume fractions of each of the condensed phases, $\{\vec \phi^{(\alpha)}\}$, indexed by $\alpha = 1,\ldots,K$.
For example, a design problem involving $N=5$ components and $K=4$ condensed phases is illustrated in \mbox{\figref{fig:1}A--B}.
\figref{fig:1}C shows the designed interaction matrix that we obtain using our method, which, along with a corresponding chemical potential vector, results in coexistence among the prescribed phases.
In general, we assume that each target condensed phase consists of $M^{(\alpha)}$ ``enriched'' components, which comprise the majority of the total non-solvent volume fraction of phase $\alpha$, and $N-M^{(\alpha)}$ ``depleted'' components, which are found at much lower concentrations in phase $\alpha$.
Such a distinction can always be made by comparing the concentration of a component within the $\alpha$ phase to its concentration in the dilute phase; if the ratio of these quantities, or \textit{partition coefficient}, for a component is greater than unity, then the component is considered to be enriched in the $\alpha$ phase.

In this work, we compute phase equilibria among bulk phases (i.e., in the thermodynamic limit), meaning that the contributions of interfacial effects to the total free energy of the fluid are negligible.
Bulk phase coexistence occurs when all $K+1$ phases have equal osmotic pressures and each molecular species has the same chemical potential in each of the $K+1$ phases.
All $K+1$ phases must also be stable with respect to concentration fluctuations, such that $\partial\vec\mu(\vec\phi)/\partial\vec\phi$ is positive definite.
Even for pairwise mean-field models of the form given by \eqref{eq:muex}, these thermodynamic conditions result in a system of nonlinear equations.
Thus, solving for $\bm\epsilon$ and $\vec\mu$ given a prescribed set of condensed phases, $\{\vec\phi^{(\alpha)}\,\forall\alpha=1,\ldots,K\}$, is a numerically challenging problem.

To find $(\bm{\epsilon},\vec\mu)$ solutions that satisfy these thermodynamic constraints in an efficient manner, we perform a \textit{convex relaxation} that relies on two assumptions about the design problem.
First, we assume that the concentration of each depleted component within each condensed phase is very small.
This assumption is equivalent to specifying the inverse-design problem in terms of only the enriched-component concentrations in each of the condensed phases.
We therefore ignore the contributions of the depleted components to the pairwise interaction term in \eqref{eq:muex} when it is evaluated in a condensed phase.
This assumption also implies that the equal-chemical-potential constraint for each depleted component $j$ in each condensed phase $\alpha$ can be approximated as an inequality, such that $\phi_j^{(\alpha)} < \phi_{\text{depl}}^{(\alpha)} \equiv \phiT^{(\alpha)}/M^{(\alpha)}(N-M^{(\alpha)})$, where the total volume fraction of non-solvent components in phase $\alpha$ is $\phi_{\text{T}}^{(\alpha)} \equiv \sum_{i=1}^N \phi_i^{(\alpha)}$.
Second, we assume that every component is enriched in at least one condensed phase and that $\phiT^{(\alpha)} \gg \phiT^{(0)}$ for every condensed phase $\alpha$, where $\phiT^{(0)}$ is the total volume fraction in the dilute phase.
This assumption implies that the osmotic pressure is approximately zero at coexistence.

Relaxing the thermodynamic constraints in these ways alters the design problem, such that $(\bm{\epsilon},\vec\mu)$ solutions to the convex relaxation may not produce the target phase diagram precisely (\figref{fig:2}A).
Nonetheless, we emphasize that this convex relaxation is only used for identifying \textit{candidate solutions} to the design problem.
As we discuss below, we subsequently verify whether a proposed solution solves the actual design problem, meaning that we achieve numerically precise phase coexistence among $K+1$ phases and that the $K$ condensed-phase compositions closely resemble those prescribed by the target phase diagram.
The degree to which the approximations described above are valid for a particular design problem only affects the ability of our method to suggest a useful candidate solution, which depends on the overlap of the actual and relaxed solution spaces (see \secref{sec:validation}).

Taken together, the design problem and the relaxed thermodynamic conditions define a semidefinite program (SDP)~\cite{boyd2004convex} that is convex with respect to the pairwise interaction matrix $\bm{\epsilon}$ and the chemical potential vector $\vec\mu$:
\begin{subequations}
  \label{eq:constraints}
  \begin{align}
    \label{eq:constraint-eqmu}
    \mu_{\text{id},i}(\vec\phi^{(\alpha)}; \vec{v}) + \mu_{\text{ex},i}(\vec\phi^{(\alpha)}; \bm{\epsilon}, \vec{v}) &\ge \mu_i \;\forall i,\alpha \\
    \label{eq:constraint-eqP}
    P(\vec\phi^{(\alpha)}; \bm\epsilon, \vec{v}) &= 0 \;\forall \alpha \\
    \label{eq:constraint-possdef}
    \partial [\vec\mu_{\text{id}}(\vec\phi^{(\alpha)}; \vec{v}) + \vec\mu_{\text{ex}}(\vec\phi^{(\alpha)}; \bm{\epsilon}, \vec{v})] / \partial \vec\phi &\succ \lambda_{\text{min}}I \;\forall \alpha \\
    \label{eq:constraint-crit}
    \phiT^{(0)}(\vec\mu;\vec v) &< \phiT^*(\vec v),
  \end{align}
\end{subequations}
where $\mu_{\text{id},i} = v_i^{-1}\log\phi_i^{(\alpha)}$ for any component $i$ that is enriched in the $\alpha$ phase and $\mu_{\text{id},i} = v_i^{-1}\log\phi_{\text{depl}}^{(\alpha)}$ for any component $i$ that is depleted in the $\alpha$ phase.
The equality(inequality) in \eqref{eq:constraint-eqmu} applies to enriched(depleted) components, respectively.
\eqref{eq:constraint-eqP} is a statement of the zero-osmotic-pressure approximation.
In \eqref{eq:constraint-possdef}, the parameter $\lambda_{\text{min}} \ge 0$ places a lower bound on the smallest eigenvalue of the second-derivative matrix to guarantee thermodynamic stability.
The final constraint, \eqref{eq:constraint-crit}, ensures that the volume fraction in the dilute phase, $\phiT^{(0)}$, is less than the critical volume fraction, $\phiT^*(\vec{v})$; this condition is independent of $\bm\epsilon$ due to the zero-osmotic-pressure assumption.
While not strictly needed when the true coexistence pressure is actually close to zero, \eqref{eq:constraint-crit} constrains $\vec\mu$ to physically plausible solutions for arbitrary phase-diagram design problems.
This convex program is straightforward to solve using modern convex optimization tools~\cite{diamond2016cvxpy,odonoghue2016conic}.
Moreover, it is possible to prove whether this convex relaxation is infeasible~\cite{boyd2004convex}, meaning that no solution $(\bm\epsilon,\vec\mu)$ exists for the target condensed-phase volume fractions $\{\vec\phi^{(\alpha)}\}$.
Finally, as long as the assumptions underlying the convex relaxation are appropriate for a proposed design problem, we expect that there will be a close correspondence between the feasible domain of $\{\vec\phi^{(\alpha)}\}$ and the domain of sets of target phases on which true thermodynamic coexistence can be established (\figref{fig:2}A).

In general, the SDP specified by \eqref{eq:constraints} defines a continuous space of interaction matrices that solve the convex relaxation of the inverse-design problem, with a unique $\vec\mu$ corresponding to each point in this space.
However, we have not yet considered the possibility that other ``off-target'' condensed phases may be equally or even more stable than the target phases at the coexistence point specified by $(\bm\epsilon,\vec\mu)$, meaning that the target phases are only in marginal or metastable coexistence.
More precisely, satisfying the equal-chemical-potential and equal-osmotic-pressure conditions for phase coexistence implies that the dilute and condensed target phases all have precisely the same value of the grand potential, $\Omega(\vec\phi;\vec\mu,\bm\epsilon,\vec{v}) \equiv \sum_{i=1}^N \int d\phi_i [v_i^{-1}\log\phi_i + \mu_{\text{ex},i}(\vec\phi) - \mu_i]$.
It is nonetheless possible that an off-target phase has a lower grand potential at the coexistence point specified by $\vec\mu$ and is thus thermodynamically favored relative to the coexisting phases specified in the design problem.
We address this possibility by introducing a regularization heuristic that attempts to maximize $\Omega(\phi;\vec\mu,\bm\epsilon,\vec{v})$ away from the target phases (\figref{fig:2}B).
Specifically, based on the form of \eqref{eq:muex}, we seek to minimize both the variance of the elements of $\vec\mu$ and the norm of $\bm\epsilon - \bm{\bar\mu} / \langle\phiT^{(\alpha)}\rangle$, where $\bar\mu_{ij} \equiv (\mu_i+\mu_j)/2$ and $\langle\phiT^{(\alpha)}\rangle$ is the mean condensed-phase total volume fraction (see \appref{app:FH} for details).
Regularizing the SDP in this way tends to destabilize off-target phases while also guaranteeing that the solution to our convex relaxation is unique.
We therefore use this heuristic to choose a particular ``regularized'' candidate solution $(\bm\epsilon,\vec\mu)$, as illustrated in \figref{fig:2}A.

Finally, to confirm that the precise thermodynamic conditions for bulk-phase coexistence are indeed satisfied by a candidate interaction matrix, we perform a multicomponent generalization of the common tangent construction.
Starting from the SDP solution $(\bm\epsilon,\vec\mu)$, we adjust $\vec\mu$ in order to fit a common tangent plane to the local minima of the grand potential, $\Omega(\vec\phi;\vec\mu,\bm\epsilon,\vec{v})$ (\figref{fig:2}C).
The conditions specified in \eqref{eq:constraints} imply that the grand potential evaluated at the SDP solution has local minima close to the prescribed condensed-phase and coexisting dilute-phase volume fractions.
Therefore, we can fit a common tangent plane by minimizing the norm of $\{\Delta\Omega^{(\alpha)}(\vec\mu)\}$, where $\Delta\Omega^{(\alpha)}(\vec\mu)$ is the difference between $\Omega(\vec\phi;\vec\mu)$ evaluated at the local minimum near the dilute phase and at the local minimum near the $\alpha$ condensed phase (see \appref{app:nonlinear-coex}).
This procedure converges rapidly using standard numerical methods~\cite{more2006levenberg}, since the convex relaxation is constructed to be a good approximation of this nonlinear hyperplane-fitting problem.
In the extensive numerical tests described below, we indeed find that a solution to the convex relaxation typically implies that the conditions for coexistence can be satisfied for the target phases to numerical precision.

\subsection{Validation of inverse-design strategy using the Flory--Huggins model}
\label{sec:validation}

This algorithm provides a scalable means of predicting whether prescribed target phases can be in simultaneous thermodynamic coexistence and, if so, for determining a coexistence point $(\bm\epsilon,\vec\mu)$.
To validate our approach, we apply this algorithm to a Flory--Huggins model~\cite{colby2003polymer} of a multicomponent polymer solution.
(See \appref{app:FH} for the corresponding SDP definitions.)
In this model, excluded volume interactions between monomers are captured by the term $\vec\mu_{\text{v}}(\vec\phi;L) = -\log(1 - \phiT) - (1 - 1/L)$ in \eqref{eq:muex}, where the molecular volume is proportional to the degree of polymerization, $L$.
In what follows, we assume that the degree of polymerization is the same for every component, and we perform calculations with $L$ ranging from $1$ to $100$.
For simplicity, we choose the same total volume fraction, $\phiT^{\alpha} = \phiT^{(\text{cond})}\!$, for each condensed phase.

\begin{figure}
  \includegraphics[width=\columnwidth]{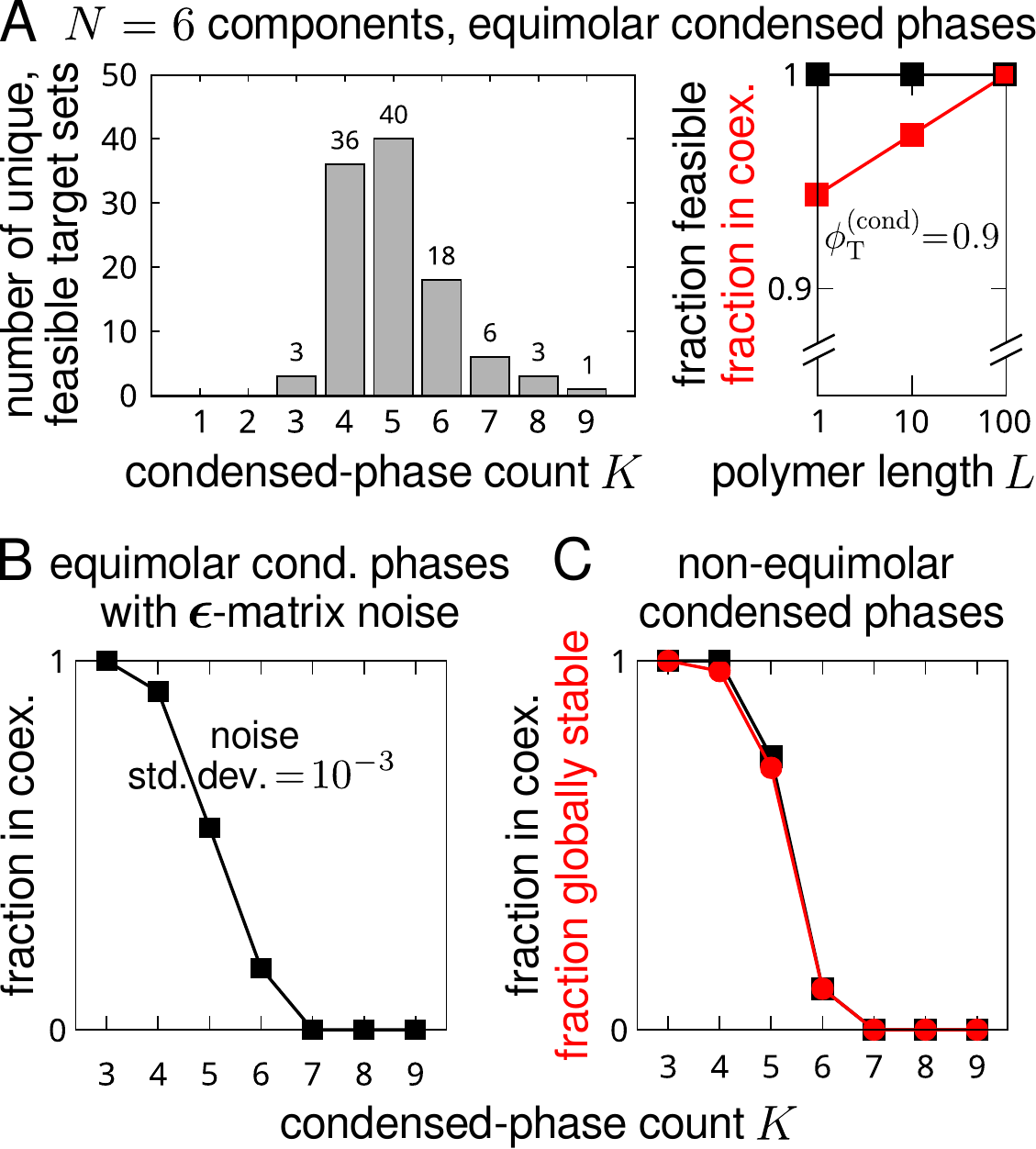}
  \caption{\textbf{Validation of the inverse-design strategy using a diverse collection of 6-component phase diagrams and the Flory--Huggins polymer model.}
    (A)~Here we test our inverse-design approach on 107 unique test cases with $N=6$ distinct non-solvent components and condensed-phase counts ranging from $K=3$ to $K=9$ (\textit{left}; see text for details).  When the target phase diagrams have equimolar condensed-phase compositions, we find that the feasibility of the convex relaxation is independent of the degree of polymerization, $L$ (\textit{right}, black).  The fraction of these convex-relaxation solutions that result in coexistence approaches one as $L$ increases (\textit{right}, red).
    (B)~Sensitivity of phase coexistence to perturbations in the interaction matrix, $\bm\epsilon$.  We add zero-mean Gaussian noise to the designed interaction matrices and then attempt to re-establish coexistence by performing a common tangent plane construction.  The probability that coexistence among all $K+1$ phases in the original target phase diagram can be re-established, averaged over many independent trials, decreases with the condensed-phase count.
    (C)~The probability that coexistence can be achieved with non-equimolar condensed-phase compositions (black).  Starting from the feasible equimolar phase diagrams, we construct target phase diagrams by randomly scaling the enriched-component compositions in each condensed phase.  We also show the probability that the coexistence point leads to global phase coexistence, meaning that no off-target phases are stable (red).  In panels C and D, $\phiT^{(\text{cond})}\! = 0.95$ and $L=100$.}
  \label{fig:3}
\end{figure}

In \figref{fig:3}, we report the results of our algorithm for a diverse collection of distinct phase diagrams with $N=6$ species.
This number of components turns out to be sufficient to uncover qualitative differences with simple fluids (see \secsref{sec:unusual} and \ref{sec:compositions}) while still permitting exhaustive searches for off-target phases.
To build a diverse collection of test cases, we begin by considering phase diagrams with ``equimolar'' target phases, meaning that every enriched component $i$ within a target phase has the same volume fraction, $\phi^{(\alpha)}_i \simeq \phiT^{(\text{cond})} / M^{(\alpha)}$.
We first enumerate phase diagrams consisting of $K$ distinct equimolar target phases (see \appref{app:generation} for details).
To eliminate trivial test cases, we require that every component is enriched in at least one condensed phase and that no two components are enriched in precisely the same set of condensed phases.
Next, we identify all phase diagrams for which the corresponding convex relaxation is feasible, using the Flory--Huggins model with degree of polymerization $L=1$.
We then group the target phase diagrams into isomorphic sets, within which phase diagrams are equivalent under permutation of component and target-phase indices.
By selecting a single target phase diagram from each isomorphic set, we end up with 107 unique $N=6$ test cases, with condensed-phase counts ranging from $K=3$ to $K=9$ (\figref{fig:3}A).
Each of these test-case phase diagrams is considered exactly once in all the calculations that follow.

Using the test cases shown in \figref{fig:3}A, we find that the feasibility of the SDP for any particular target phase diagram is independent of both $L$ and $\phiT^{(\text{cond})}\!$.
This observation suggests that the feasibility of the SDP does not depend on the contribution from $\mu_{\text{v}}$ in \eqref{eq:muex}.
However, the probability that a solution to the regularized SDP results in phase coexistence (to machine precision, $\Delta\Omega \sim 10^{-14}$) tends to increase with both $L$ and $\phiT^{(\text{cond})}$ (\figref{fig:3}A).
These trends can be explained by noting that the total dilute-phase volume fraction $\phiT^{(0)} \rightarrow 0$ as $L \rightarrow \infty$ and $\phiT^{(\text{cond})} \rightarrow 1$ in the Flory--Huggins model~\cite{colby2003polymer}.
Under these conditions, the dilute phase is nearly ideal.
The coexistence pressure therefore tends to zero, and the convex relaxation becomes a more accurate approximation of the true inverse-design problem, as $L \rightarrow \infty$ and $\phiT^{(\text{cond})} \rightarrow 1$.
Nonetheless, our algorithm succeeds in establishing coexistence among the target phases for the vast majority of our test cases even when the zero-pressure approximation is poor (e.g., with $L=1$ and $\phiT^{(\text{cond})} = 0.9$ in \figref{fig:3}A).

These equimolar-condensed-phase test cases demonstrate that our approach is capable of identifying solutions to a variety of phase-diagram design problems.
We emphasize that these design problems are non-trivial, since individual components can be enriched in multiple immiscible phases.
Furthermore, this algorithm can just as easily be applied to phase-diagram design problems with arbitrary condensed-phase volume fractions, as opposed to equimolar targets, as we discuss below.
Finally, we note that this approach is computationally efficient, returning SDP solutions with $N=6$ components in less than a second on a single core (see \appref{app:computation-time} for further discussion of the computational requirements).

\subsection{Unusual phase coexistence in mixtures with five or more non-solvent components}
\label{sec:unusual}

Our calculations reveal a number of unexpected features of multicomponent phase coexistence.
Most strikingly, we identify numerically precise coexistence points ($\Delta\Omega \sim 10^{-14}$) where the condensed-phase count, $K$, is greater than the number of distinct non-solvent species, $N$ (\figref{fig:3}A).
At first glance, these examples might appear to conflict with the Gibbs Phase Rule (GPR), which relates the number of coexisting phases to the number of thermodynamic degrees of freedom~\cite{gibbs1878equilibrium}.
In the case of the incompressible $N$-component fluids that we study here, the bound implied by a standard interpretation of the GPR is $K \le N$.
However, our results indicate that this bound does not apply to all possible coexistence points when $N=6$ (\figref{fig:3}A).

The resolution to this paradox is that, because the interactions are free parameters, it is possible to design phase-coexistence conditions that are \textit{linearly dependent}.
This linear dependence makes it possible to perform a common tangent plane construction even when $K > N$.
We emphasize that this scenario does not violate a rigorous derivation of the GPR that counts only linearly independent thermodynamic constraints.
A further consequence of linearly dependent coexistence conditions is that the lever rule~\cite{porter2021phase}, which relates the total concentrations in a mixture to the mole fractions of the coexisting phases, does not have a unique solution when $K > N$.

The origin of these unusual coexistence points can be most easily understood by realizing that the design problem, with $N(N+1)/2$ tunable interaction-matrix parameters, is not necessarily overdetermined when $K > N$.
Thus, convex optimization is able to identify interaction matrices that result in linearly dependent coexistence equations, as required to have $K > N$ condensed phases.
Our calculations indicate that these unusual coexistence conditions can only occur in mixtures with at least $N=5$ species, and that they become increasingly common as the number of components increases.
In fact, our inverse-design approach shows that it is possible to find coexistence points with more than $N^2$ condensed phases.
(See \appref{app:graph-theory} for further discussion and scaling predictions based on graph-theoretical arguments.)

If these unusual coexistence points are allowed by linearly dependent coexistence conditions, then we should expect that they are sensitive to small changes in the interaction parameters.
Consistent with this expectation, we find that small, random perturbations to the designed interaction matrices preclude phase coexistence of the target phases when $K > N$ (\figref{fig:3}B).
Specifically, we add zero-mean Gaussian noise to the designed matrix $\bm\epsilon$, and then attempt to perform a common tangent plane construction for phases close to the original target phases by tuning $\vec\mu$.
Whenever $K > N$, only a subset of the original $K$ condensed phases can be brought into coexistence with the dilute phase after such a perturbation, while the remaining condensed phases become metastable.

\subsection{Designing mixtures with arbitrary condensed-phase compositions}
\label{sec:compositions}

\vskip-0.5ex
Next, we turn our attention to variations in the compositions of the condensed phases.
Intuition based on the phase behavior of simple mixtures suggests that small changes in the target-phase volume fractions, $\{\vec\phi^{(\alpha)}\}$, should result in small changes in $\bm\epsilon$, and vice versa, unless the mixture is near a critical point where two or more of the $\vec\phi^{(\alpha)}$ merge.
For example, small changes in the dimensionless interaction parameter in an incompressible binary mixture perturb the binodal but do not change the coexistence region qualitatively, as long as $\phiT \gg \phiT^*$~\cite{colby2003polymer}.
Since the calculations presented in \figref{fig:3} are performed far from critical points (meaning that the Euclidean distance between all pairs of target phases $\{\vec\phi^{(\alpha)}\}$ is large and $\phiT^{\text{(cond)}} \gg \phiT^{(0)}$), one might expect that this intuition should apply to mixtures with many components as well.

To test this hypothesis, we randomly perturb the initially equimolar compositions of the enriched components in each target phase and then apply our inverse-design algorithm.
For most phase diagrams with $K < N$, we find that the convex relaxation with non-equimolar condensed phases is feasible and that phase coexistence can be established by a common tangent plane construction (black curve in \figref{fig:3}C).
Furthermore, exhaustive sampling of the grand potential landscapes confirms that the target phases of these designed phase diagrams are almost always globally stable (red curve in \figref{fig:3}C).
This observation provides evidence that our regularization heuristic (see \secref{sec:inverse-design}) is working as intended.

Yet in other cases, the convex relaxation becomes infeasible when non-equimolar compositions are prescribed, indicating that phase coexistence is not always possible with arbitrary condensed-phase compositions.
In particular, we find that random condensed-phase compositions are always infeasible when $K > N$.
We stress that this observation does not imply that $K > N$ coexistence points necessarily \textit{require} equimolar condensed-phase compositions.
Rather, linearly dependent coexistence conditions can still be achieved with non-equimolar condensed-phase compositions, but these compositions cannot be chosen randomly.
(See \appref{app:extended} for further analysis.)
Our calculations also show that random composition perturbations can render many phase diagrams infeasible when $K \le N$ as well.

Taken together, \figsref{fig:3}B and \ref{fig:3}C suggest that unusual coexistence points, which are sensitive to small perturbations in $\bm\epsilon$, lie on manifolds of lower dimension than the full space of interaction matrices, or ``$\bm\epsilon$-space''.
Phase coexistence is not limited to $K \le N$ condensed phases on these special manifolds, although some of these phases must become metastable if we move off the manifold by perturbing the interaction matrix.
These manifolds represent ``interfaces'' between volumes of $\bm\epsilon$-space that correspond to condensed phases with different sets of enriched components.
In other words, crossing one of these interfaces by changing $\bm\epsilon$ entails a discontinuous transition from one set of condensed phases to another, where phases from both sets are stable on the interface itself.

\subsection{Relationships between phase diagrams in the space of pairwise interactions}
\label{sec:relationships}

To probe relationships between distinct multicomponent phase diagrams in $\bm\epsilon$-space, we can analyze dissimilarities between pairs of interaction matrices that solve different phase-diagram design problems.
For this purpose, we use the Frobenius norm, $||\bm\epsilon_s - \bm\epsilon_r||_{\text{fro}}$, to measure the ``distance'' in $\bm\epsilon$-space between two interaction matrices $\bm\epsilon_r$ and $\bm\epsilon_s$, which correspond to different phase diagrams with globally stable condensed phases $\{\vec\phi^{(\alpha)}\}_r$ and $\{\vec\phi^{(\alpha)}\}_s$, respectively.
We can then represent the $N(N+1)/2$-dimensional $\bm\epsilon$-space in two dimensions via dimensionality reduction techniques, which preserve the relative distances between the interaction matrices.

\begin{figure}
  \includegraphics[width=\columnwidth]{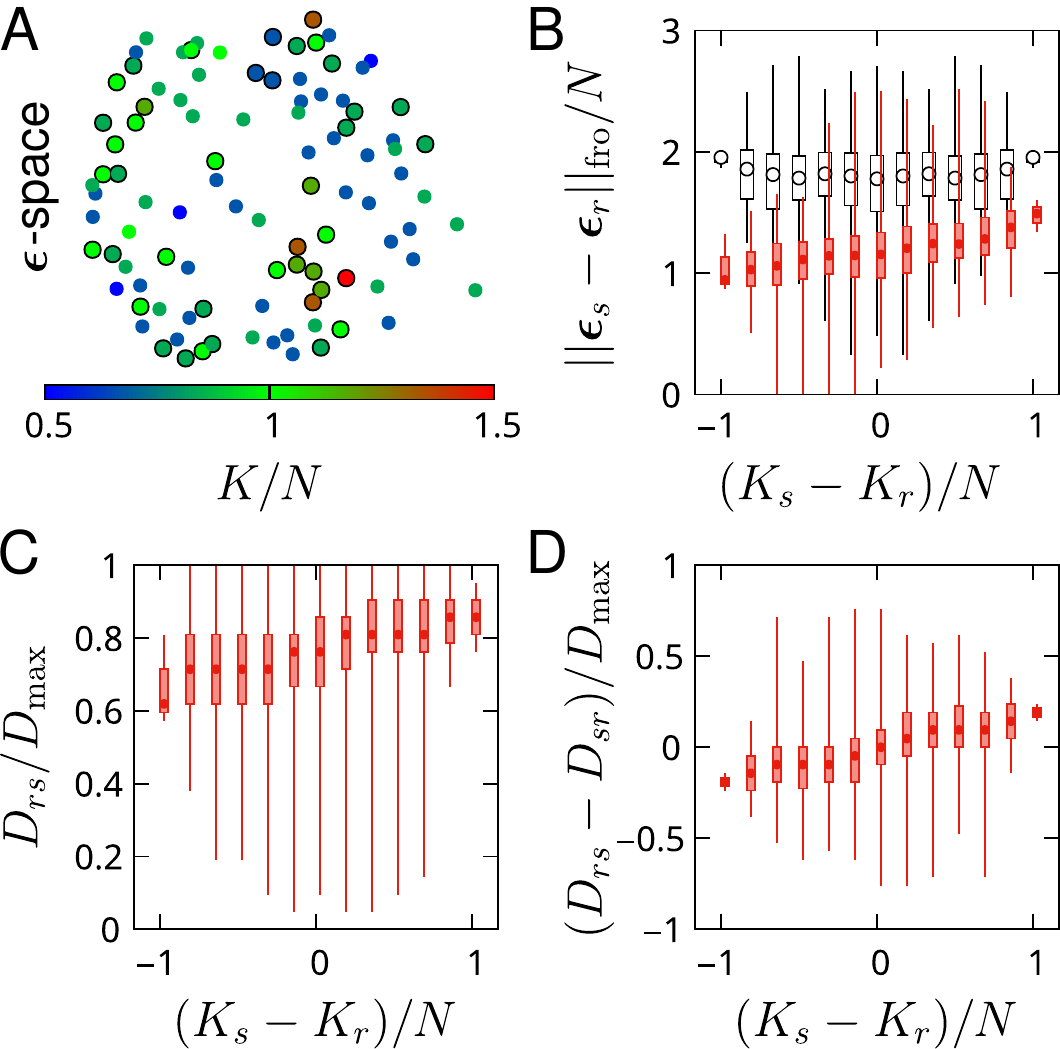}
  \caption{\textbf{Relationships among phase diagrams in $\bm\epsilon$-space.}
    (A)~A low-dimensional representation of the interaction matrices corresponding to the phase diagrams considered in \figref{fig:3}A.  Circles with black outlines indicate phase diagrams that are sensitive to random perturbations in $\bm\epsilon$ (see text and \figref{fig:3}B).  Multidimensional scaling~\cite{kruskal1964multidimensional} has been used to preserve distances in $\bm{\epsilon}$-space, taken here as the Frobenius norm of the difference between each pair of interaction matrices.
    (B)~Distributions of the distances between pairs of matrices in panel A (black distributions) are compared to distributions of the minimum distance required to switch from an initial reference phase diagram $r$ to a new phase diagram $s$ (red distributions).  Distributions are shown as a function of the condensed-phase-count difference, $K_s - K_r$.  Box plots indicate the quartiles of the distance distributions.
    (C)~Distributions of the minimum number of entries of the symmetric $\bm{\epsilon}$-matrix that must be changed to switch from a reference phase diagram $r$ to a new phase diagram $s$, $D_{rs}$.  The maximum number of elements that can be changed is $D_{\text{max}} \equiv N(N+1)/2$.
    (D)~The minimum number of elements changed when switching from a reference phase diagram $r$ to the new phase diagram $s$ is asymmetric with respect to the condensed-phase-count difference, $K_s - K_r$.}
  \label{fig:4}
\end{figure}

A two-dimensional representation of the interaction-matrix solutions to the equimolar design problems presented in \figref{fig:3}A is shown in \figref{fig:4}A.
On the basis of this projection, we conclude that the similarity between a pair of interaction matrices is not directly related to how many condensed phases they encode.
To see this more clearly, we plot the distribution of distances between these pairs of interaction matrices as a function of the difference in their condensed-phase counts, $K_s - K_r$ (black distributions in \figref{fig:4}B).
In this way, we find that the typical distance between pairs of interaction matrices tends to be relatively constant regardless of whether they encode the same number of condensed phases.

However, because the interaction matrix that stabilizes a particular phase diagram is typically not unique, it is more useful to quantify the extent to which an interaction matrix must be changed in order to switch to a different phase diagram.
We can address this question within our inverse-design framework by modifying the regularization heuristic in one of two ways (see \appref{app:alt-regularization} for details).
In the first instance, we attempt to minimize the distance to a reference interaction matrix that solves a different phase-diagram design problem.
For example, starting from the interaction matrix $\bm\epsilon_r$ that solves the original regularized SDP for phase diagram $r$, we can identify the ``closest'' matrix $\bm\epsilon_s$ that solves phase diagram $s$ by minimizing the Frobenius norm $||\bm\epsilon_s - \bm\epsilon_r||_{\text{fro}}$ (red distributions in \figref{fig:4}b).
In this way, we find that increasing the number of condensed phases, such that $K_s > K_r$, tends to require more substantial changes in the interaction matrix as measured by this distance metric.
Interestingly, this distance can in fact be infinitesimal if the phase-diagram change $r \rightarrow s$ reduces the phase count; this occurs whenever the initial interaction matrix $\bm\epsilon_r$ resides on a special low-dimensional manifold (see \secsref{sec:unusual} and \ref{sec:compositions}) where $K > N$ and the two phase diagrams $r$ and $s$ differ by a single condensed phase.

Alternatively, we can determine the smallest number of distinct matrix elements that must be changed in order to switch to a new phase diagram.
As should be expected, this minimal number of elementwise changes, $D_{rs}$, is always greater than zero, even when the initial interaction matrix resides on a special low-dimensional manifold with $K > N$ (\figref{fig:4}C).
Our calculations reveal that $D_{rs}$ is asymmetric with respect to phase diagram changes $r \leftrightarrow s$ and tends to increase with the net number of added phases (\figref{fig:4}D).
Assuming that the reference interaction matrix is obtained using the original regularization heuristic, this observation implies that a larger number of distinct matrix elements typically need to be modified when adding, as opposed to subtracting, condensed phases.

\subsection{Validation of inverse-design strategy in a molecular simulation model}
\label{sec:simulation}

Finally, we assess whether the predictions of our inverse-design approach apply beyond mean-field models.
We therefore investigate whether we can design the phase behavior of non-mean-field fluid models in which the molecules interact via \textit{pair potentials}, meaning that the contribution to the total potential energy from every pair of molecules depends on the distance between them~\cite{hansen2013theory}.
In models with pair potentials, the chemical potentials cannot be described exactly by the mean-field pairwise approximation that has been central to our discussion up to this point.
In particular, \eqref{eq:muex} is only a good approximation of the excess chemical potentials at low concentrations, since the higher-order virial coefficients depend on the species-specific pair potentials~\cite{hansen2013theory}.

To this end, we perform simulations and free-energy calculations to compute phase coexistence using a multicomponent lattice gas.
We first design a mean-field interaction matrix, $\bm\epsilon^{\text{MF}}$, for a target phase diagram using the $L=1$ Flory--Huggins SDP.
We then use this matrix to define the pair potentials, $u_{ij}$, on a three-dimensional square lattice with lattice constant $a$.
Each lattice site can be occupied by at most one solute molecule at a time, so that $u_{ij}(r/a < 1) = \infty$, where $r$ is the distance between solutes of types $i$ and $j$.
We set the well-depths for interactions between neighboring molecules to be proportional to the designed mean-field interaction matrix, such that $u_{ij}(1 \le r/a < 2) \propto \epsilon_{ij}^{\text{MF}}$.
We first identify the free-energy basins, which correspond to the (meta)stable phases of the lattice gas, by running grand-canonical Monte Carlo simulations~\cite{frenkel2001understanding}.
We then sample reversible transitions between the dilute free-energy basin and each of the condensed-phase basins~\cite{jacobs2013predicting}.
Finally, we reconstruct the free-energy landscapes in the $N$-dimensional $\vec\phi$-space and adjust the chemical potentials to bring all phases into coexistence~\cite{shirts2008statistically}, at which point the grand potentials of all basins are all equal (see \appref{app:fe-calc} for details).

\begin{figure}
  \includegraphics[width=\columnwidth]{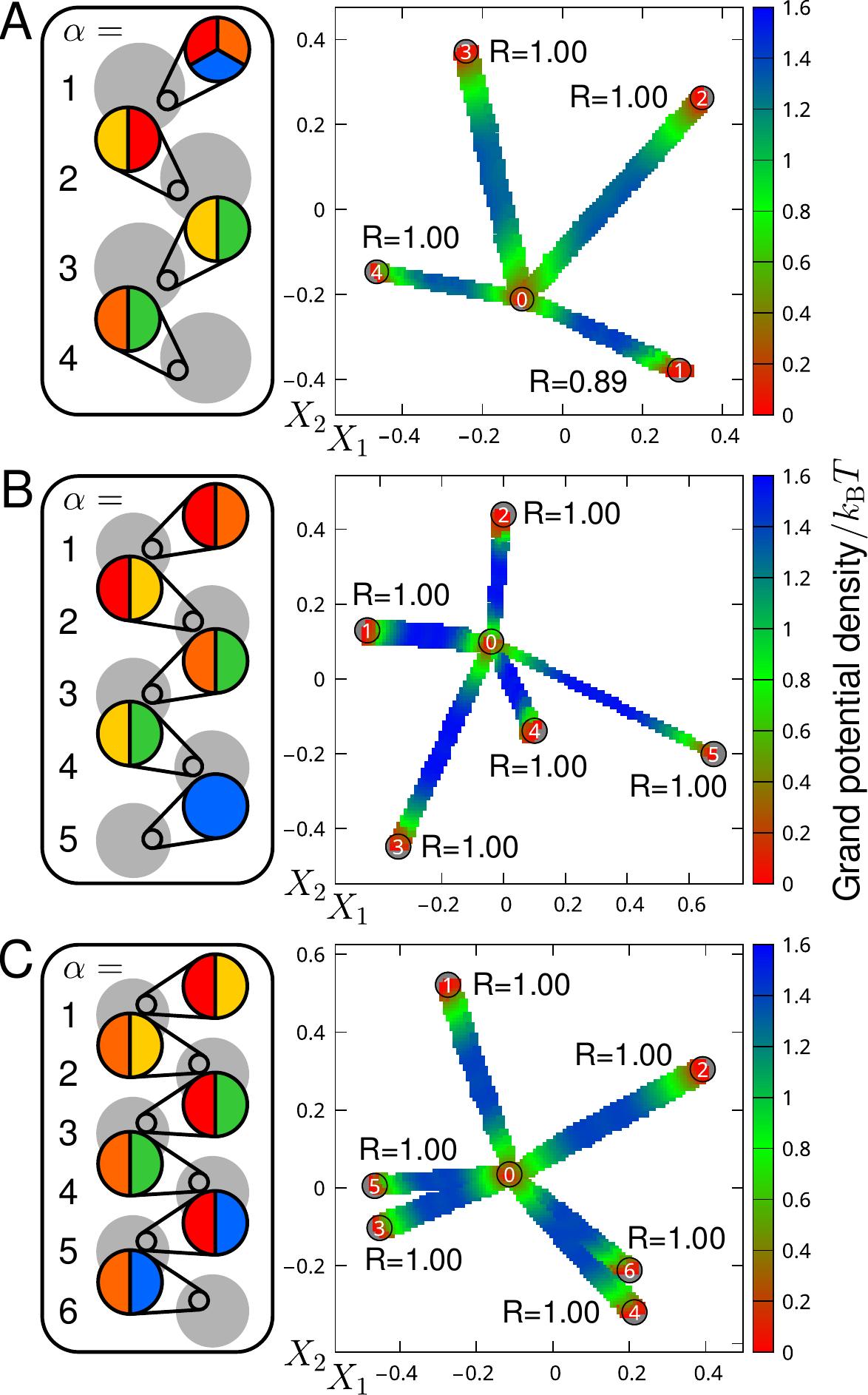}
  \caption{\textbf{Transferable predictions validate the pairwise approximation.}  We perform free-energy calculations using a multicomponent lattice-gas model and pair potentials derived from designed interaction matrices.  Reweighting techniques are then used to compute phase coexistence to within sampling accuracy ($|\Delta\Omega| \le 0.007 k_{\text{B}}T$).  Examples are shown for mixtures with $N=5$ species and (A)~$K=4$, (B)~$5$, and (C)~$6$ condensed phases.  The $N$-dimensional concentration spaces are projected onto two coordinates, $X_1$ and $X_2$, for visualization in two dimensions.  The value of the grand potential, which is determined along a path connecting each condensed phase to the dilute phase, is indicated by the color scale.  Each labeled phase on the projected landscape corresponds to a condensed phase in the schematic to the left.  The Pearson correlation coefficient, R, between the target and simulated composition is shown for each condensed phase.}
  \label{fig:5}
\end{figure}

Our lattice-gas simulations reveal free-energy landscapes that are consistent with the target phase diagrams, even though the pair potentials are designed using mean-field interaction matrices.
In order to test a variety of scenarios, we carry out simulations with five components and condensed-phase counts that are less than, equal to, and greater than the number of components.
In \figref{fig:5}, projections of the high-dimensional grand-potential landscapes are visualized in two dimensions for each of these test cases.
These landscapes indicate that phase coexistence is achieved to within statistical error ($|\Delta\Omega| \le 0.007 k_{\text{B}}T$) among all the prescribed phases.
In particular, we find that phase coexistence is in fact achieved in the $K=6$ example, confirming that unusual phase coexistence with $K>N$ condensed phases is not only a feature of mean-field models.
We note that minor quantitative differences in the phase compositions do occur in the simulation model, however (\figref{fig:5}).
These inaccuracies arise due to the mean-field approximations utilized in the interaction-matrix design algorithm and appear to become more significant as the number of enriched components within a condensed phase grows.
Nonetheless, the identities of the enriched components, if not their precise target compositions, match the designs in all the simulated phases in each test case.

\section{Conclusions}

Understanding how biomolecular interactions give rise to \textit{many} coexisting phases with distinct chemical compositions remains an outstanding problem with significant implications for intracellular biophysics.
To address this challenge, we have introduced an \textit{inverse} approach for designing mixtures that demix into phases with prescribed compositions.
This methodology provides insight into the structure of the interaction-matrix solution space, revealing a number of qualitative differences between multicomponent phase behavior and that of simple fluids.
Our approach also establishes an extensible framework for investigating relationships between the physicochemical properties of biomolecules and phase separation in complex mixtures.
For example, we could build on the approach described in \secref{sec:relationships} to compute \textit{pairs} of interaction matrices that enable switching between different phase behaviors with the fewest number of matrix-element changes.
Our approach could similarly be applied to design interfaces~\cite{mao2020designing,chew2023thermodynamic} between condensed phases.

Our inverse-design approach differs in a number of important ways from existing numerical methods for analyzing phase coexistence in multicomponent fluids~\cite{jacobs2017phase,shrinivas2021phase,zwicker2022evolved}.
First, by employing a convex relaxation of thermodynamic constraints, convex optimization can be used to prove whether a proposed phase diagram is infeasible~\cite{jacobs2021self}.
This feature allows us to distinguish between the consequences of actual physical constraints and the behavior of approximate and/or non-deterministic algorithms for computing or designing phase coexistence.
Second, inverse design avoids the need to search high-dimensional spaces to identify candidate phases when solving the forward problem.
In the context of iterative design schemes, such as those based on genetic algorithms, such calculations must be performed for every candidate interaction matrix.
Third, inverse design is well suited for solving highly nontrivial design problems.
For example, we have shown that we can design fluids by prescribing not only the \textit{number} of coexisting condensed phases, but their precise \textit{compositions} as well.
This distinction becomes important in experimentally relevant scenarios where he total molecular concentrations are fixed.
In this case, constraints on the compositions of the condensed phases must be specified in the design problem to guarantee that the total concentration vector lies within the coexistence region, ensuring that the mixture phase separates into the intended number of coexisting phases (e.g., \figref{fig:1}B).

Going beyond mean-field models, we anticipate that our approach might be applied to design the phase behavior of more realistic multicomponent fluids.
Supporting this idea, our lattice-model simulations---in which the interactions can be described as only \textit{approximately} pairwise---suggest that the predictions of our approach may indeed translate to more complex systems.
However, it may not always be possible to engineer or evolve molecular interactions with the independence and precision necessary to construct all theoretically possible phase diagrams.
In this regard, our results indicate that the physically relevant constraints on the phase behavior of multicomponent fluids arise primarily from the properties of the intermolecular interactions, since thermodynamically allowed phase diagrams can be surprisingly complex.
Our results therefore highlight the need to understand the extent to which molecular interactions can be tuned independently in phase-separating (bio)chemical fluids.
Addressing this challenge will require the introduction of additional constraints on the physicochemical properties of the molecular components within our design framework.
In this way, we anticipate that our theoretical approach will play an important role in ongoing efforts to unravel the connections between molecular design and multicomponent phase behavior~\cite{simon2017programming,espinosa2020liquid,heidenreich2020designer,sanders2020competing,kaur2021sequence,do2022engineering,baruch2023biomolecular,lyons2023functional}.

Source code and example calculations are available at \href{https://github.com/wmjac/phaseprogramming-pub}{\texttt{https://github.com/wmjac/phaseprogramming-pub}}.

This work is supported by the National Science Foundation (DMR-2143670).

\appendix
\renewcommand\thefigure{A\arabic{figure}}    
\setcounter{figure}{0}

\section{Application to the multicomponent Flory--Huggins model}
\label{app:FH}

In the numerical examples presented in \secsref{sec:validation}--\ref{sec:relationships}, we consider the special case of the multicomponent Flory--Huggins polymer model~\cite{colby2003polymer}, for which
\begin{equation}
  \vec\mu_{\text{v}}(\vec\phi;\vec{v}) = -\log(1 - \phiT) - (1 - 1/L_i),
\end{equation}
and we introduce the degree of polymerization, $L_i$, in place of the molecular volume $v_i$.
In this section, we first define the thermodynamic quantities for this model explicitly.
We then formulate the corresponding SDP.

\subsection{Model definition}

The Helmholtz free-energy density, $F$; chemical potential, $\vec\mu$; osmotic pressure, $P$; and Hessian matrix, $\partial\vec\mu(\vec\phi) / \partial\vec\phi$, in the multicomponent Flory--Huggins model are
\begin{align}
  F &= \sum_{i=1}^N \frac{\phi_i}{L_i} \log \phi_i + (1 - \phiT) \log (1 - \phiT) \nonumber \\
  &\qquad + \frac{1}{2} \sum_{i=1}^N\sum_{j=1}^N \epsilon_{ij} \phi_i \phi_j \\
  \mu_i &= \frac{1}{L_i} \log \phi_i - \log (1 - \phiT) - \left(1 - \frac{1}{L_i}\right) + \sum_{j=1}^N \epsilon_{ij} \phi_j \\
  P &= -\log (1 - \phiT) + \sum_{i=1}^N \frac{\phi_i}{L_i} - \phiT + \frac{1}{2} \sum_{i=1}^N\sum_{j=1}^N \epsilon_{ij} \phi_i \phi_j \\
  \frac{\partial \mu_i}{\partial \phi_j} &= \frac{\delta_{ij}}{L_i\phi_i} + \frac{1}{1 - \phiT} + \epsilon_{ij},
\end{align}
respectively, where $L_i$ is the degree of polymerization of polymeric species $i$.

Before writing down the SDP constraints for a particular set of condensed-phase volume fractions, $\{\vec\phi^{(\alpha)}\}$, we consider a mixture with a fixed composition $\vec x$.
The resulting expressions will be utilized in subsequent sections.
The mixture composition is normalized such that $\sum_{i=1}^N x_i = 1$.
In order to calculate the total volume fraction at the critical point, $\phiT^*(\vec x)$, we set the projection of the Hessian matrix along $\vec x$ to zero,
\begin{align}
  &\sum_{i=1}^N\sum_{j=1}^N x_i \frac{\partial\mu_i}{\partial\phi_j} x_j \nonumber \\
  &\qquad = \frac{1}{\phiT} \sum_{i=1}^N \frac{x_i}{L_i} + \frac{1}{1 - \phiT} + \sum_{i=1}^N\sum_{j=1}^N x_i \epsilon_{ij} x_j = 0,
\end{align}
and differentiate with respect to $\phiT$,
\begin{equation}
  -\frac{\partial\langle\epsilon\rangle_x}{\partial\phiT} = -\frac{1}{(\phiT)^2} \sum_{i=1}^N\frac{x_i}{L_i} + \frac{1}{(1 - \phiT)^2} = 0,
\end{equation}
where $\langle \epsilon \rangle_x \equiv \sum_{i=1}^N\sum_{j=1}^N x_i \epsilon_{ij} x_j$.
The critical volume fraction at fixed composition $\vec x$ is thus
\begin{equation}
  \label{eq:phiT-crit}
  \phiT^*(\vec x) = \frac{1}{1 + \langle1 / L\rangle_x^{-1/2}},
\end{equation}
where $\langle 1 / L \rangle_x \equiv \sum_{i=1}^N x_i / L_i$.

Assuming that the chemical potential vector is known, the total volume fraction of a condensed phase with composition $\vec x$ can be approximated by setting the osmotic pressure equal to zero and projecting $\vec\mu$ along $\vec x$,
\begin{widetext}
\begin{equation}
  \sum_{i=1}^N x_i\mu_i \equiv \langle\mu\rangle_x = \sum_{i=1}^N \frac{x_i}{L_i}\log\phi_i + \langle 1 / L\rangle_x - 1 - \log(1 - \phiT) + \phiT \langle\epsilon\rangle_x,
\end{equation}
to yield a non-linear equation for $\phiT$,
\begin{equation}
  \frac{2}{\phiT}\log(1 - \phiT) + \langle 1 / L \rangle\log\phiT - \log(1 - \phiT) + 1 - \langle 1 / L \rangle_x + \sum_{i=1}^N \frac{x_i}{L_i}\log x_i - \langle\mu\rangle_x = 0.
\end{equation}
\end{widetext}
We can also solve for the mean interaction, $\langle\epsilon\rangle_x$, in a condensed phase with composition $\vec x$,
\begin{equation}
  \label{eq:eps-mean}
  \langle \epsilon \rangle_x = \frac{2}{(\phiT)^2} \left[\log(1 - \phiT) + \phiT \left(1 - \langle 1 / L \rangle_x \right)\right].
\end{equation}
We can similarly obtain an expression for dilute-phase volume fractions in terms of $\vec\mu$ by assuming that the osmotic pressure is nearly zero, such that $\phiT^{(0)}$ is very small,
\begin{equation}
  \label{eq:phi-dilute}
  \phi_i \simeq \exp\left[L_i\mu_i + (L_i - 1)\right].
\end{equation}

\subsection{SDP formulation}
\label{sisec:SDP}

In this section, we use the notation $\{\vec\phi^{(\alpha)}\}$ to refer to the target volume fractions in the $K$ condensed phases, $\alpha = 1, \ldots, K$.
We further assume that $M^{(\alpha)}$ species are enriched in the $\alpha$ phase and that the target volume fractions of the depleted components are set to zero.
We can therefore denote the set of enriched components in phase $\alpha$ by
\begin{equation}
  \label{eq:target-set}
  S^{(\alpha)} \equiv \{ i | \delta(\phi_i^{(\alpha)}) = 0 \},
\end{equation}
where $\delta(\cdot)$ is the Dirac delta function.
Thus, ${M^{(\alpha)} = N - \sum_{i=1}^N \delta(\phi_i^{(\alpha)})}$ is the cardinality of the vector $\vec\phi^{(\alpha)}$, and the target total volume fraction in phase $\alpha$ is ${\phiT^{(\alpha)} = \sum_{i\in S^{(\alpha)}} \phi_i^{(\alpha)}}$.

The equal chemical potential constraints for enriched and depleted components, respectively, take the form
\begin{widetext}
\begin{subequations}
  \label{eq:equal-mu}
  \begin{align}
    \sum_{j=1}^N L_i \phi_j^{(\alpha)} \epsilon_{ij} - L_i \mu_i + \log\phi_i^{(\alpha)} - L_i \log(1 - \phiT^{(\alpha)}) - (L_i - 1) &= 0, \quad \delta\left(\phi_i^{(\alpha)}\right) = 0 \\
    \sum_{j=1}^N L_i \phi_j^{(\alpha)} \epsilon_{ij} - L_i \mu_i + \log\left[\frac{\phiT^{(\alpha)} \zeta}{M^{(\alpha)} (N - M^{(\alpha)})} \right] - L_i \log(1 - \phiT^{(\alpha)}) - (L_i - 1) &\ge 0, \quad \delta\left(\phi_i^{(\alpha)}\right) = 1
  \end{align}
\end{subequations}
\end{widetext}
for each species index $i = 1, \ldots, N$ and each condensed phase $\alpha$.
We set the adjustable parameter $\zeta = 10^{-2}$ in this work.
The zero-osmotic-pressure constraint for each condensed phase $\alpha$ is
\begin{equation}
  \label{eq:zero-pressure}
  \frac{1}{2} \sum_{i=1}^N\sum_{j=1}^N \phi_i^{(\alpha)} \phi_j^{(\alpha)} \epsilon_{ij} - \log(1 - \phiT^{(\alpha)}) - \phiT^{(\alpha)} + \sum_{i=1}^N \frac{\phi_i^{(\alpha)}}{L_i} = 0.
\end{equation}
In order to place constraints on the Hessian matrices in the condensed phases, we define the regularized target volume fractions
\begin{equation}
  \tilde\phi_i^{(\alpha)} = \phi_i^{(\alpha)} + \frac{\phiT^{(\alpha)} \zeta}{M^{(\alpha)} (N - M^{(\alpha)})} \delta\left(\phi_i^{(\alpha)}\right),
\end{equation}
so that we have $\tilde\phi_i^{(\alpha)} > 0$ for all $i$ and all $\alpha$.
Each condensed-phase scaled Hessian matrix must then satisfy
\begin{equation}
  \label{eq:hessian}
  \frac{\delta_{ij}}{\tilde\phi_i^{(\alpha)}} + \frac{\sqrt{L_i L_j}}{1 - \phiT^{(\alpha)}} + \sqrt{L_i L_j} \epsilon_{ij} \succeq \lambda_{\text{min}} \delta_{ij},
\end{equation}
where $\lambda_{\text{min}}$ is the smallest allowed eigenvalue.
We choose $\lambda_{\text{min}} = 1$ in this work.
Finally, we constrain the chemical potentials such that the approximate total volume fraction of a roughly equimolar dilute phase is below the critical volume fraction by utilizing \eqref{eq:phi-dilute} and \eqref{eq:phiT-crit},
\begin{align}
  &\log \sum_{i=1}^N \exp(L_i\mu_i + L_i - 1) \nonumber \\
  &\qquad \le \log \left\{ 0.9\, \phiT^*\left[\vec x = \left(\frac{1}{N}, \ldots, \frac{1}{N}\right)\right] \right\}.
\end{align}
Within the approximations of this convex relaxation, these constraints define the joint space of interaction matrices, $\bm{\epsilon}$, and chemical potential vectors, $\vec\mu$, for which bulk phase coexistence can be established among the target condensed phases and a dilute phase.

\subsection{Regularization for global stability of target phases}
\label{sisec:regularization-stability}

Next, we regularize our convex optimization problem in order to identify an interaction matrix, $\bm\epsilon$, and chemical potential vector, $\vec\mu$, that are least likely to result in stable off-target phases (\figref{fig:2}A).
Off-target condensed phases correspond to local minima of the grand-potential, $\Omega(\vec\phi;\vec\mu,\bm\epsilon,\vec{v})$, that lie below the grand potential of the dilute phase, $\Omega^{(0)}$.
We therefore aim to maximize the grand potential everywhere in the domain $\vec\phi$, except at the dilute and target phases, $\{\vec\phi^{(0)}, \vec\phi^{(1)}, \ldots, \vec\phi^{(K)}\}$, where $\Omega^{(\alpha)} = \Omega^{(0)}$.
Since the $\vec\mu$ and $\bm\epsilon$-dependent contributions to the grand potential have the form $\omega(\vec\phi) \equiv \left(\vec\phi^\top\bm\epsilon - \vec\mu\right) \cdot \vec\phi$, we define the objective function
\begin{equation}
  \label{eq:L0}
  \mathcal{L}_0 = \Bigg|\!\Bigg|(1 + \delta_{ij})^{1/2} \frac{\bar\omega_{ij}}{\langle\phiT\rangle_{\{\alpha\}}}\Bigg|\!\Bigg|_{\text{fro}} + \Bigg|\!\Bigg|L_i \left(\mu_i - \frac{1}{N} \sum_{k=1}^N \mu_k\right)\!\Bigg|\!\Bigg|_2.
\end{equation}
The first term is a scaled and shifted Euclidean norm of the unique $\bar\omega_{ij}$ elements, 
\begin{equation}
  \frac{\bar\omega_{ij}}{\langle\phiT\rangle_{\{\alpha\}}} = \epsilon_{ij} - \frac{1}{\langle\phiT\rangle_{\{\alpha\}}} \left(\frac{\mu_i + \mu_j}{2} - \frac{1}{N} \sum_{k=1}^N \mu_k\right),
\end{equation}
where $\langle\phiT\rangle_{\{\alpha\}} \equiv (1/K) \sum_{\alpha=1}^K \phiT^{(\alpha)}\!$, while the second term is a scaled standard deviation of the $\vec\mu$ elements.
The notations $||\cdot||_{\text{fro}}$ and $||\cdot||_2$ indicate the matrix Frobenius norm and vector Euclidean norm, respectively.

In the calculations presented in \secsref{sec:validation}--\ref{sec:compositions}, we solve an SDP in which we minimize $\mathcal{L}_0$ while obeying the constraints described in \appref{sisec:SDP}.
In order to suppress off-target condensed phases that are enriched in a single component, we also introduce an additional constraint on the on-diagonal elements of $\bm\epsilon$,
\begin{equation}
  \min_{i \in S^{(\alpha)}} \epsilon_{ii} \ge \langle\epsilon\rangle_{x^{(\alpha)}},
\end{equation}
where $\langle\epsilon\rangle_{x^{(\alpha)}}$ is defined in \eqref{eq:eps-mean}.
As shown in \figref{fig:3}C, this regularization heuristic has the intended effect of biasing the SDP solution towards coexistence points for which no off-target phases are stable.

\subsection{Regularization for minimum interaction-matrix dissimilarity}
\label{app:alt-regularization}

In \figref{fig:4}B--D, we illustrate how interaction matrices must be changed in order to switch from a reference phase diagram to a new phase diagram while making minimal modifications to the interaction matrix.
Let us assume that $\bm\epsilon_r$ is the interaction matrix that solves the original SDP, regularized by \eqref{eq:L0}, for a set of target phases $\{\vec\phi^{(\alpha)}\}_r$.
To identify the interaction matrix $\bm\epsilon_s$ that is ``closest'' to this given initial matrix $\bm\epsilon_r$ while satisfying phase coexistence among a different set of target phases $\{\vec\phi^{(\alpha)}\}_s$, we define a new objective function
\begin{align}
  \mathcal{L}_{\text{d}}(w) &= w \Bigg|\!\Bigg|(1 + \delta_{ij})^{1/2} \frac{\bar\omega_{ij}}{\langle\phiT\rangle_{\{\alpha\}}}\Bigg|\!\Bigg|_{\text{fro}} \nonumber \\
  &\qquad + \Bigg|\!\Bigg|L_i \left(\mu_i - \frac{1}{N} \sum_{k=1}^N \mu_k\right)\!\Bigg|\!\Bigg|_2 + ||\bm\epsilon_s - \bm\epsilon_r||,
\end{align}
where $w \ge 0$ is an adjustable parameter.
When minimizing the distance between $\bm\epsilon_s$ and $\bm\epsilon_r$ (see \figref{fig:4}B), we use the Frobenius norm for the final term in $\mathcal{L}_{\text{d}}$.
When attempting to minimize the number of distinct elements of the symmetric matrix $\bm\epsilon_r$ that must be changed in order to establish coexistence among the target phases $\{\vec\phi^{(\alpha)}\}_s$ (see \figref{fig:4}C--D), we make use of the convex relaxation of vector cardinality; consequently, we use the L1 norm of the upper triangle of $\bm\epsilon_s - \bm\epsilon_r$ as the final term in $\mathcal{L}_{\text{d}}$ in this case.
In both cases, we find the smallest value of the adjustable parameter, $w^*$, for which $\bm\epsilon_s$ results in the target phase diagram $s$ with no off-target phases.
This calculation is carried out to a precision of $w^* \pm 10^{-3}$.
The distributions shown in \figref{fig:4}B--D are computed using $\mathcal{L}_{\text{d}}(w^*)$, with $w^*$ calculated independently for each phase diagram change $r \rightarrow s$.

\section{Numerical verification of phase coexistence in mean-field models}
\label{app:exact-coex}

Given a solution to the SDP, we identify the exact coexistence point for the target condensed phases $\{\vec\phi^{(\alpha)}\}$, if it exists, by solving the non-linear coexistence equations to numerical precision.
In this section, we describe the numerical procedures that we follow to establish bulk phase coexistence and to check for stable off-target phases.

\subsection{Non-linear phase coexistence solver}
\label{app:nonlinear-coex}

The grand potential, $\Omega(\vec\phi)$, of a mean-field model with an excess chemical potential in the form of \eqref{eq:muex} can be written as
\begin{widetext}
\begin{equation}
  \Omega(\vec\phi;\vec\mu,\bm\epsilon,\vec{v}) \equiv \sum_{i=1}^N \int \!d\phi_i \left[\frac{\log\phi_i}{v_i} + \mu_{\text{ex},i}(\vec\phi) - \mu_i\right]
  = \sum_{i=1}^N \left[ \frac{\phi_i (\log \phi_i - 1)}{v_i} + \int \!d\phi_i\, \mu_{\text{v}}(\vec\phi) + \frac{1}{2} \sum_{j=1}^N \epsilon_{ij}\phi_i\phi_j - \mu_i\phi_i \right].
\end{equation}
\end{widetext}
In order to find a hyperplane that is co-tangent to the local minima of the grand potential corresponding to the dilute and target condensed phases, we define the grand potential difference for each condensed phase~${\alpha = 1,\ldots,K}$,
\begin{equation}
  \Delta\Omega^{(\alpha)}(\vec\mu; \bm\epsilon, \vec{v}) \equiv \Omega(\phi_{\text{sp}}^{(\alpha)}; \vec\mu,\bm\epsilon,\vec{v}) - \Omega(\phi_{\text{sp}}^{(0)}; \vec\mu,\bm\epsilon,\vec{v}),
\end{equation}
where $\Omega(\phi_{\text{sp}}^{(\alpha)})$ and $\Omega(\phi_{\text{sp}}^{(0)})$ indicate the grand potential evaluated at the stationary point (i.e., the local minimum) of $\Omega(\vec\phi;\vec\mu,\bm\epsilon,\vec{v})$ nearest to phase $\alpha$ or the dilute phase, respectively.
In practice, we identify these stationary points by minimizing $\Omega(\vec\phi;\vec\mu,\bm\epsilon,\vec{v})$, starting from either a target condensed-phase volume fraction or from the approximate dilute-phase volume fraction, \eqref{eq:phi-dilute}, using the Newton conjugate gradient trust-region algorithm~\cite{nocedal2006conjugate}.
We then minimize the Euclidean norm of the $K$-dimensional $\vec\Delta\Omega(\vec\mu)$ vector by iteratively updating the chemical potential vector $\vec\mu$ and locating the stationary points to calculate $\Delta\Omega^{(\alpha)}(\vec\mu; \bm\epsilon, \vec{v})$ for each condensed phase.
Minimization of this Euclidean norm is carried out using the Levenberg--Marquardt nonlinear least squares (NLLS) algorithm~\cite{more2006levenberg}.
The conditions for bulk-phase coexistence are satisfied when this norm reaches machine precision ($\lesssim 10^{-14}$).

\subsection{Identification of stable off-target condensed phases}
\label{app:off-target}

We can perform a brute-force search for off-target stable phases by minimizing the grand potential at coexistence, starting from randomly generated initial points in $\vec\phi$-space.
The grand potential at coexistence, $\Omega(\vec\phi;\vec\mu)$, is first determined via the nonlinear phase coexistence solver described above, which fixes $\vec\mu$.
To perform one trial of the search, we generate an initial point $\vec\phi_{\text{trial}}$ by sampling uniformly from the $N$-dimensional unit simplex, such that $\phi_{\text{trial},i} > 0$ for all components ${i = 1,\ldots,N}$ and $\sum_{i=1}^N \phi_{\text{trial},i} < 1$.
We then use the Newton conjugate gradient trust-region algorithm~\cite{nocedal2006conjugate} to minimize $\Omega(\vec\phi)$ starting from this initial point.
This algorithm terminates upon reaching a local minimum, $\vec\phi_{\text{sp}}$, on the grand-potential surface.
If $\vec\phi_{\text{sp}}$ differs from the dilute, $\vec\phi_{\text{sp}}^{(0)}$, and target-phase, $\vec\phi_{\text{sp}}^{(\alpha)}$, local minima, then we compare the grand potential evaluated at this new local minimum, $\Omega_{\text{sp}} = \Omega(\vec\phi_{\text{sp}};\vec\mu)$, to the coexistence grand potential, $\Omega(\vec\phi_{\text{sp}}^{(0)};\vec\mu)$.
A new local minimum is deemed to be a stable off-target phase if $\Omega_{\text{sp}} \le \Omega(\vec\phi_{\text{sp}}^{(0)})$ to within a numerical tolerance of $10^{-3}$.

In \secsref{sec:compositions} and \ref{sec:relationships}, we carry out $10^4$ trials in order to determine whether any off-target stable phases exist at a proposed coexistence point $(\bm\epsilon,\vec\mu)$.
We find that this number of trials is sufficient to yield consistent, reproducible results for mixtures with $N \le 6$.
However, we emphasize that this approach is computationally expensive, since each trial involves an $N$-dimensional minimization, and the number of trials must scale exponentially with $N$ in order to carry out a sufficiently exhaustive search.
Thus, while we can use this algorithm to validate the results of our inverse-design regularization heuristic for small $N$, any direct usage of this brute-force approach (or any brute-force approach, for that matter) would not be scalable to mixtures with a much larger number of components.

\section{Generation of target phase compositions}
\label{app:generation}

\subsection{Enumeration of target sets}
\label{app:target-sets}

In an effort to explore a wide variety phase diagrams, we enumerate ``target sets'' with a fixed number of components in \secref{sec:validation}.
As in \eqref{eq:target-set}, target sets label each of the $N$ components as being either enriched or depleted in each of the $K$ condensed phases.
We shall therefore refer to target sets as specifying the ``topology'' of the phase diagram.
We enumerate all phase-diagram topologies by generating $(K,N)$ target sets that satisfy the following conditions:
\begin{enumerate}
\item Each of the $N$ components is enriched in at least one of the $K$ targets, and
\item No two components are enriched in precisely the same targets.
\end{enumerate}
The second condition prevents consideration of phase diagrams that have fewer than $N$ independent components.
In other words, if two components were to be enriched in precisely the same target phases, then the rows and columns of $\bm\epsilon$ associated with these components would also be directly related; therefore, a phase diagram with smaller $N$ but an equivalent topology could be constructed by grouping these components together.
Note that these conditions result in a finite lower bound on $K$.
For example, these conditions cannot be satisfied using $N=6$ components if $K<3$.

Two $(K,N)$ targets sets are isomorphic if they can be made identical by permuting the ordering of the components and/or target phases.
To account for this, we sort all target sets into isomorphic groups and consider one member of each group in all calculations.
Specifically, we choose a target set within an isomorphic group by sorting the target phases in decreasing order of the enriched component cardinality, $M^{(\alpha)}$, and the components in decreasing order of the number of target phases in which each component appears.
Sorting in this way allows us to compute the minimum $\bm\epsilon$-space distance and the minimum number of changed $\bm\epsilon$ elements between target phase diagrams (\figref{fig:4}B--D).

In the results presented in \secsref{sec:validation}--\ref{sec:relationships}, we generate ``equimolar'' target phases by choosing the target volume fractions in the condensed phases to be $\phi_i^{(\alpha)} = \phiT^{(\text{cond})} / M^{(\alpha)}$ if component $i$ is enriched in phase $\alpha$ and $\phi_i^{(\alpha)} = 0$ otherwise.
We then generate ``non-equimolar'' target phases by randomly scaling the volume fractions of the enriched components in each phase of an equimolar target set.
To this end, we define a scale factor, $s > 0$, and scale each volume fraction by $\phi_i^{(\alpha)} \leftarrow \phi_i^{(\alpha)} [1 + \exp(s \eta)]$, where $\eta$ is a random number between 0 and 1.
Finally, we adjust each target phase such that the total volume fraction is equal to $\phiT^{(\text{cond})}$ using the transformation $\phi_i^{(\alpha)} \leftarrow (\phiT^{(\text{cond})} / \phiT^{(\alpha)}) \phi_i^{(\alpha)}$.

\subsection{Scaling of the maximum phase count based on graph theory}
\label{app:graph-theory}

In \refcite{jacobs2021self}, we showed that the feasibility of a related convex optimization problem can be predicted on the basis of graph-theoretical arguments.
Specifically, we showed that the problem of designing a mean-field free-energy landscape with prescribed local minima reduces to a quadratic program (QP) if all the condensed phases are enriched in exactly the same number of components, $M$, and the composition of each target phase is equimolar.
Under these special conditions, the feasibility of the QP can be predicted by considering the maximal cliques~\cite{west2001introduction} within a graph, $G$, as follows.
The vertices of $G$ correspond to the $N$ species, and the adjacency matrix is defined according to
\begin{equation}
  G_{ij} = \begin{cases}
    1 & \text{if components }i\text{ and }j\text{ are both enriched} \\
    \, & \qquad\text{in any phase }\alpha \\
    0 & \text{otherwise}.
  \end{cases}
\end{equation}
The components enriched in each target phase define a subset of the vertices of $G$, as noted in \eqref{eq:target-set}.
If any of these subsets are not maximal cliques in $G$, then the QP is infeasible.

Extending this argument to the present work, we propose that the feasibility of phase coexistence among $K$ equimolar condensed phases that satisfy the equal-$M$ condition described above can be predicted using the same graph-theoretical approach.
Thus, for these special cases, determining the phase count reduces to the problem of finding maximal cliques in $G$.
For example, to construct phase diagrams in which the phase count scales \textit{quadratically} with the number of components, $N$, we can enrich every condensed phase with precisely two components. This scaling follows from Tur\'{a}n's theorem~\cite{Turan}, which states that the maximum number of edges of a graph free of 3-cliques is $N^2 / 4$, in which case every edge is a maximal clique corresponding to a target phase.

We can also apply this argument to estimate the largest possible condensed-phase count in an $N$-component mixture.
Graphs with extremal clique counts can be realized by partitioning the $N$ components into subsets of size 3 (assuming $N$ is divisible by 3), and creating edges between all pairs of components that are not in the same subset~\cite{Moon}.
This construction results in target phases enriched in precisely $M=N/3$ components, while the combinatorial nature of this construction gives rise to a phase-count scaling that is \textit{exponential} with respect to $N$, $K \sim 3^{N/3}$.
We have verified that equimolar target phases generated via this construction lead to phase coexistence with the prescribed phase count.
For example, applying our inverse-design approach to equimolar target phases with $N=15$ and $M=5$ results in numerically precise phase coexistence ($\Delta\Omega < 4\times10^{-13}$) among $K=243$ condensed phases.

\section{Scaling analysis of computation time}
\label{app:computation-time}

We benchmark the computational cost of solving the convex relaxation by running our algorithm, implemented using state-of-the-art convex optimization software~\cite{diamond2016cvxpy,odonoghue2016conic}, on randomly generated target sets with an increasing number of components, $N$.
To ensure that the probability of generating a feasible target set does not go to zero as the number of components grows, we consider two sampling schemes for which the average total number of enriched components across all condensed phases, $\sum_{\alpha=1}^K M^{(\alpha)}$, scales linearly with the number of components.
In the first scheme, we sample target sets with a constant number of condensed phases, $K$.
In practice, we choose $K=6$ random integers from the domain $[1,2^N]$ uniformly without replacement, and then use the bit-string representation of each integer to define a target phase (where a 1 indicates an enriched component and a 0 indicates a depleted component in each phase $\alpha = 1,\ldots,K$).
This approach maintains a constant number of conic constraints, \eqref{eq:constraint-possdef}, in the design problem.
In the second scheme, we scale $K$ linearly with $N$, such that the number of conic constraints increases with the number of components.
In practice, we generate $K=N/2$ phases, with enriched-component cardinality $M^{(\alpha)}$ chosen uniformly on the domain $[1,5]$ for each phase $\alpha$.
We then construct the set of enriched components in each phase $\alpha$ via Bernoulli trials with probability $M^{(\alpha)} / N$.
Since the mean enriched-component cardinality $\langle M^{(\alpha)} \rangle$ is constant, the mean total number of enriched components across all phases is proportional to $K$, and thus to $N$.
In both schemes, we always ensure that there is at least one component enriched in each phase and that every phase has a unique set of enriched components. 

\begin{figure}
  \includegraphics[width=\columnwidth]{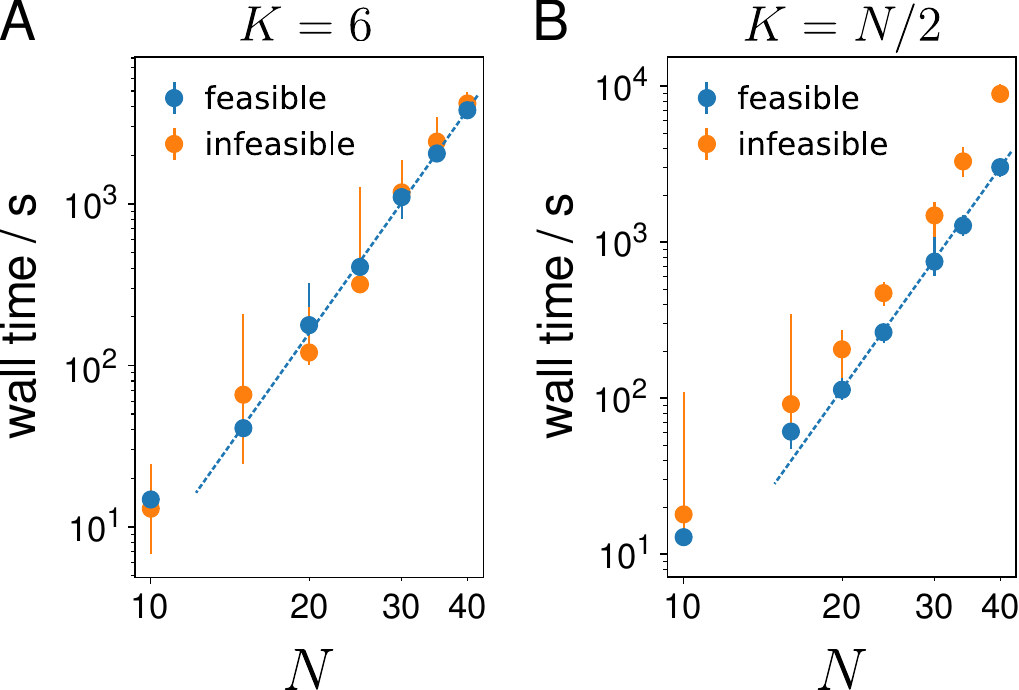}
  \caption{The median computation time required to solve the convex relaxation or to prove the infeasibility of randomly generated design problems as a function of the number of components, $N$.  (A)~In the first scheme, the number of condensed phases, $K$, is constant.  (B)~In the second scheme, the number of condensed phases grows proportionally to $N$.  Error bars are computed via bootstrapping.  Dashed lines, indicating approximate power-law scalings, are guides to the~eye.}
  \label{appfig:computation-time}
\end{figure}

\figref{appfig:computation-time}A and \figref{appfig:computation-time}B show the median computation time required to solve the design problems generated in these two ways, respectively, with error bars computed via bootstrapping.
In both schemes, the median computation time needed to obtain the regularized solution to a feasible convex relaxation is consistent with power-law scaling.
However, the median time required to prove infeasibility appears to increase exponentially with $N$ in the second scheme (\figref{appfig:computation-time}B).
In all cases, the computation time required to solve the common tangent plane construction is small by comparison (i.e., $\lesssim 10\%$ of the total computation time when $N=20$, and $\lesssim 1\%$ of the total computation time when $N=40$).

\section{Unusual features of multiphase coexistence}
\label{app:extended}

\subsection{Sensitivity analysis of designed interaction matrices}

\begin{figure*}
  \includegraphics[width=0.875\textwidth]{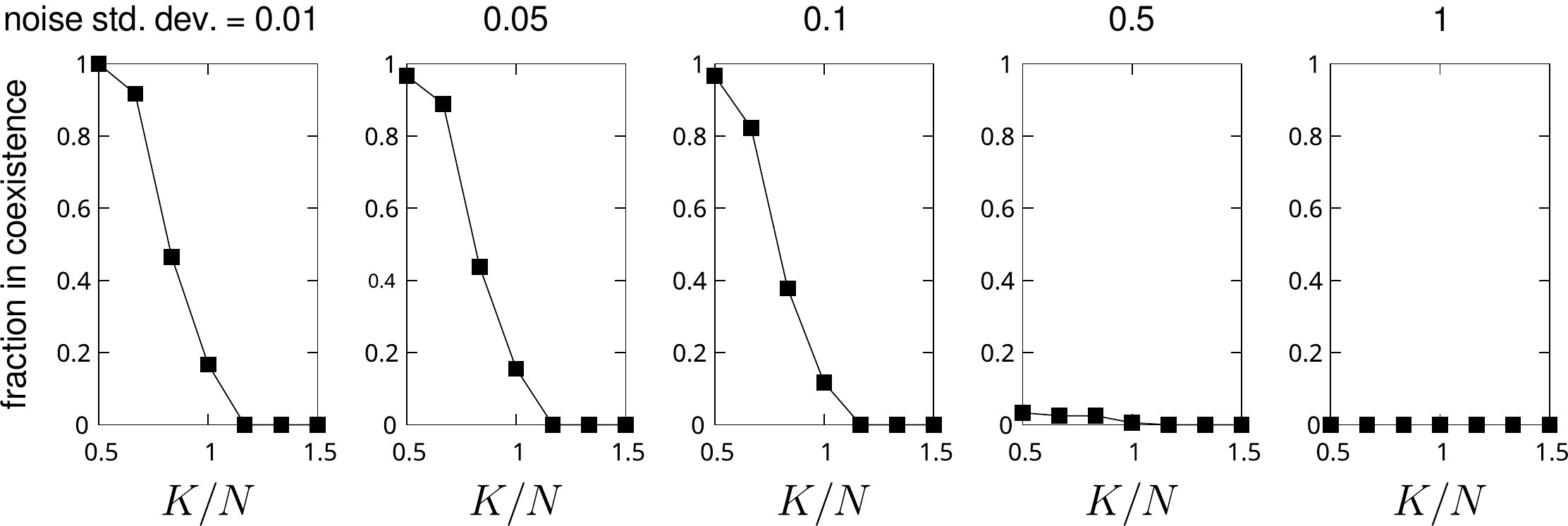}
  \caption{The probability that phase coexistence among target phases with $N=6$ species can be re-established after zero-mean Gaussian noise is added to the interaction matrix, as a function of the condensed-phase count, $K$, and the noise standard deviation.  Calculations are performed using the Flory--Huggins polymer model with $\phiT^{(\text{cond})}\! = 0.95$ and $L=100$ (cf.~\figref{fig:3}B).}
  \label{appfig:noise}
\end{figure*}

\begin{figure*}
  \includegraphics[width=0.925\textwidth]{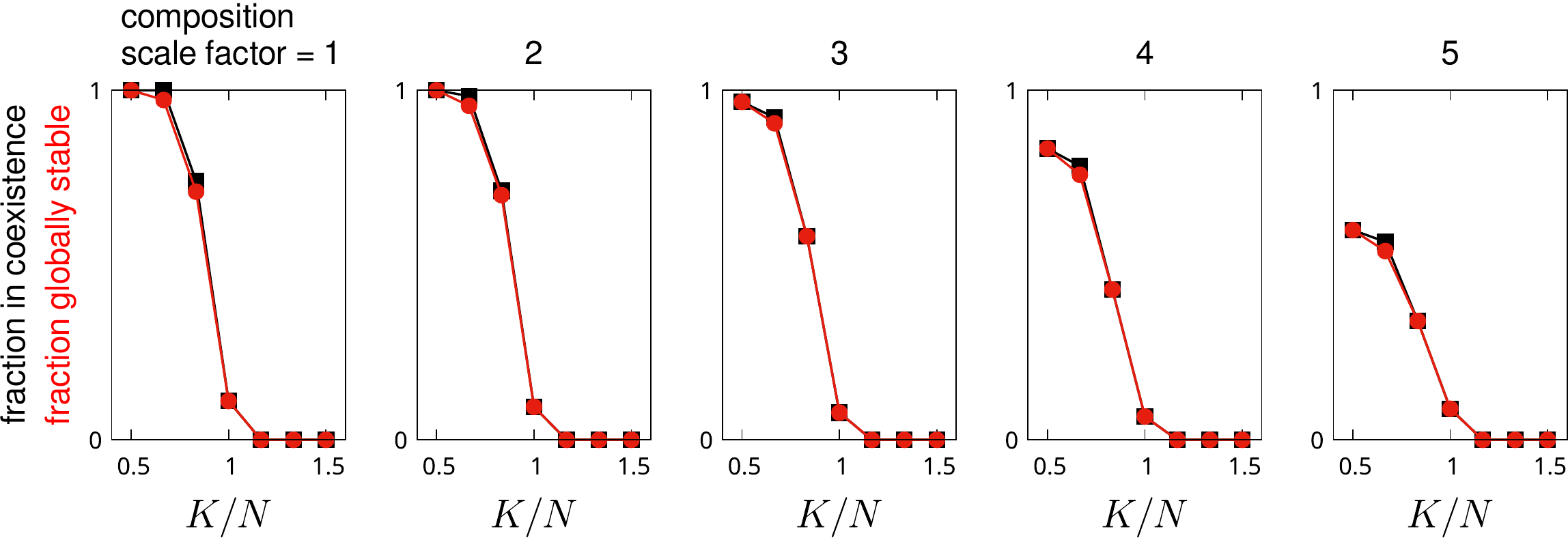}
  \caption{The probability that phase coexistence (black points) can be established for condensed phases with $N=6$ species and randomly generated non-equimolar compositions, assuming that the associated equimolar phase diagram (i.e., the phase diagram with equimolar condensed phases having the same enriched components) is feasible.  We also report the probability that the target non-equimolar phases are globally stable at the designed coexistence point (red points).  Results are shown as a function of the condensed-phase count, $K$, and the scale factor $s$ (see \appref{app:target-sets}) used to randomize the compositions of the enriched components in the condensed phases.  Calculations are performed using the Flory--Huggins polymer model with $\phiT^{(\text{cond})}\! = 0.95$ and $L=100$ (cf.~\figref{fig:3}C).}
  \label{appfig:composition}
\end{figure*}

As described in \secsref{sec:unusual} and \ref{sec:compositions}, we analyze the sensitivity of designed equimolar-target-phase coexistence points to random perturbations in both the interaction matrices and the condensed-phase volume fractions.
Extended results are shown in \figref{appfig:noise}, where we systematically vary the strength of the zero-mean Gaussian noise added to the designed interaction matrices, and in \figref{appfig:composition}, where we systematically vary the scale factor used to alter the initially equimolar enriched-component compositions in the target condensed phases (see \appref{app:target-sets}).
In both cases, increasing the magnitude of the perturbation, either by increasing the standard deviation of the Gaussian noise (\figref{appfig:noise}) or by increasing the composition scale factor (\figref{appfig:composition}), tends to reduce the probability that coexistence can be re-established among the same number of target phases.

If the perturbations in $\bm\epsilon$-space or $\vec\phi$-space are sufficiently large, then it is not possible to re-establish coexistence for any set of initial target phases.
(See, e.g., results with a noise standard deviation of 1 in \figref{appfig:noise}.)
This behavior can be understood by noting that large perturbations may cross a critical manifold, at which point the topology of the phase diagram changes and it becomes impossible to re-establish coexistence among the original $K$ target phases.
In the case of smaller perturbations (e.g., a noise standard deviation of 0.01 in \figref{appfig:noise} or a composition scale factor of 1 in \figref{appfig:composition}), random perturbations tend to destabilize one or more of the target phases in all cases when $K > N$, as well as in select cases when $K \le N$.
Nonetheless, we emphasize that the perturbations considered in \figref{appfig:noise} and \figref{appfig:composition} are \textit{random}.
In the following section, we analyze how the situation changes when the interaction matrix is perturbed in a \textit{non-random} manner.

\subsection{Analysis of compositional constraints via iterative perturbation}

\begin{figure*}
   \includegraphics[width=\textwidth]{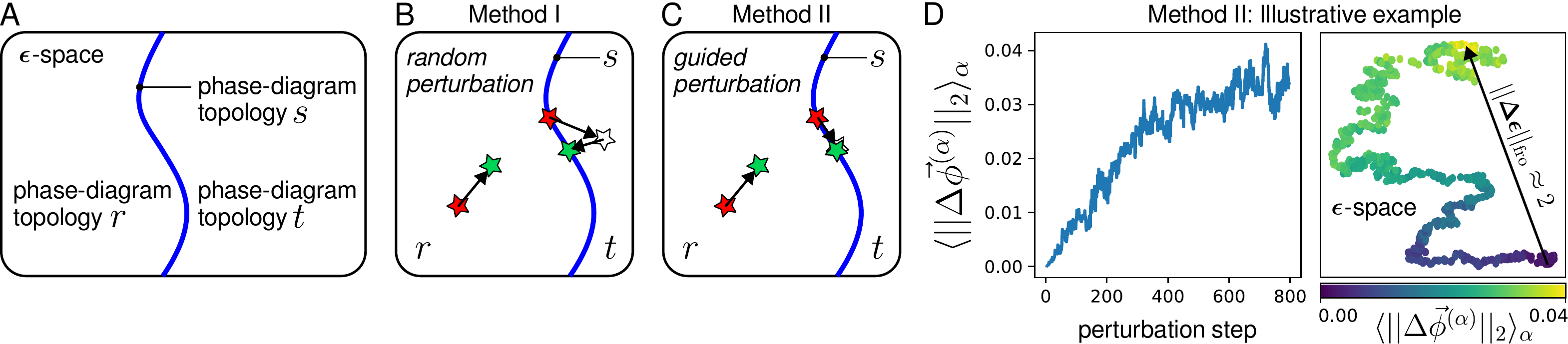}\vskip-1ex
   \caption{(A)~Schematic of subspaces, each corresponding to a different phase-diagram topology, within the full $\bm\epsilon$-space.  In this illustration, the solution space corresponding to phase-diagram topology $s$ has a lower dimension than the full $\bm\epsilon$-space.
     (B,C)~Schematics of two methods for iteratively perturbing a coexistence point while maintaining a given phase-diagram topology.  The equimolar coexistence point is indicated by a red star.  In Method I, we randomly perturb $\bm\epsilon$ (empty star), and then use NLLS minimization to re-establish phase coexistence consistent with the target phase-diagram topology (green star).  In Method II, we perturb $\bm\epsilon$ and $\vec\mu$ in accordance with linearized coexistence equations (see text).
     (D)~Repeated application of Method II results in diffusive behavior of the root-mean-squared distance between the condensed-phase volume fractions at the perturbed and initial coexistence points, averaged over all condensed phases, $\langle||\Delta\phi^{(\alpha)}||_2\rangle_{\alpha}$.  Projecting the interaction matrices at the perturbed coexistence points via multidimensional scaling~\cite{kruskal1964multidimensional} shows the path taken through $\bm\epsilon$-space.}
  \label{appfig:perturbation}
\end{figure*}

The results shown in \figref{appfig:noise} suggest that the interaction-matrix solution space, corresponding to a target phase-diagram topology, can have a lower dimension than the full $\bm\epsilon$-space.
In these cases, random perturbations to the interaction matrix destabilize one or more phases---changing the phase-diagram topology---because the perturbed coexistence point is moved off of this lower-dimensional manifold.
This scenario is illustrated schematically by phase-diagram topology $s$ in \figref{appfig:perturbation}A.
However, this picture suggests that it should be possible to maintain the target phase-diagram topology by proposing perturbations that move along the lower-dimensional manifold.
In this section, we show how this can be done, allowing us to find non-equimolar coexistence points on such low-dimensional manifolds.

One method for finding non-equimolar coexistence points makes direct use of the nonlinear coexistence solver introduced in \appref{app:nonlinear-coex}.
Specifically, we use NLLS to minimize $\vec\Delta\Omega(\vec\mu,\bm\epsilon)$, except here we allow both $\vec\mu$ \textit{and} $\bm\epsilon$ to change.
To this end, we further modify the least-squares objective function to force the NLLS solver to find a coexistence point involving all $K$ condensed phases that are present in target phase-diagram topology.
This brute-force approach (Method I, \figref{appfig:perturbation}B) generically leads to a new coexistence point among non-equimolar condensed phases.
We note that for phase-diagram topologies lying on low-dimensional manifolds in $\bm\epsilon$-space (e.g., topology $s$ in \figref{appfig:perturbation}B), the changes in the volume fractions between the equimolar and non-equimolar coexistence points tend to be correlated across target phases and components.

Alternatively, we can obtain non-equimolar phase diagrams by systematically perturbing $\vec{\mu}$ (Method II, \figref{appfig:perturbation}C).
Starting from a set of feasible equimolar target phases $\{\vec{\phi}^{(\alpha)}\}_0$ at a designed coexistence point $(\bm\epsilon_0, \vec{\mu}_0)$, we can expand $\Omega$ and $\vec{\mu}$ to linear order in $\bm{\Delta\epsilon}$ and $\Delta\vec{\phi}$,
\begin{widetext}
\begin{equation}
  \label{eq:Omega-linear}
  \Omega^{(\alpha)} = \Omega_0^{(\alpha)} + \left[\left.\frac{\partial F}{\partial \vec{\phi}}\right|_0^{(\alpha)}
    \right]^\top \!\!\cdot \Delta \vec{\phi}^{(\alpha)} + 
  \left[\left.\frac{\partial F}{\partial \vec{\epsilon}}\right|_0^{(\alpha)} 
    \right]^\top \!\!\cdot \Delta\vec{\epsilon} 
  - \vec{\mu}_0^\top \cdot \Delta\vec{\phi}^{(\alpha)} - \Delta\vec{\mu}^\top \!\cdot \vec{\phi}_0^{(\alpha)}
  = \Omega_0^{(\alpha)} + \left[\left.\frac{\partial F}{\partial \vec{\epsilon}}\right|_0^{(\alpha)}
    \right]^\top \!\!\cdot \Delta \vec{\epsilon} - \Delta\vec{\mu}^\top \!\cdot \vec{\phi}_0^{(\alpha)},
\end{equation}
\end{widetext}
where $\vec{\epsilon}$ denotes the vector containing the independent elements of $\bm\epsilon$.
Here we have assumed that  $(\bm\epsilon_0, \vec{\mu}_0)$ is located far from a critical manifold.
For phase coexistence to be maintained (to linear order) for some perturbation $\bm{\Delta\epsilon}$, $\Omega^{(\alpha)} = \Omega^{(0)}$ must hold for all condensed phases $\alpha=1,\ldots,K$.
Thus, from \eqref{eq:Omega-linear}, we obtain a system of linear equations of the form $\bm A \Delta \vec{\epsilon} = \bm B \Delta\vec{\mu} = \vec{b}$, where $\bm A\in \mathcal{R}^{K \times N(N+1)/2}$, $\bm B\in \mathcal{R}^{K \times N}$, $\Delta \vec{\epsilon} \in \mathcal{R}^{N(N+1)/2}$, and $\Delta \vec{\mu} \in \mathcal{R}^{N}$:
\begin{align}
  \left[ \left.\frac{\partial F}{\partial \vec{\epsilon}}\right|_0^{(\alpha)} - \left.\frac{\partial F}{\partial \vec{\epsilon}}\right|_0^{(0)} \right]^\top \!\!\cdot \Delta\vec{\epsilon} &= \left[ \vec{\phi}_0^{(\alpha)} - \vec{\phi}_0^{(0)} \right]^\top \!\!\cdot \Delta\mu \nonumber \\
  &\qquad\qquad\forall \alpha=1,\ldots,K. 
\end{align}
The matrices $\bm A$ and $\bm B$ may be rank deficient when the volume fractions of the equimolar target phases are linearly dependent.
When ${\text{col}(\bm A) = \text{col}(\bm B)}$, solutions are guaranteed for arbitrary perturbations, corresponding to cases in which ${\text{rank}(\bm A)=\text{rank}(\bm B)=K \leq N}$.
If $K > N$, then $\text{rank}(\bm B) < \text{rank}(\bm A) \leq N$, and random perturbations will in general fail.
However, it is still possible to find solutions for some perturbation $\Delta \vec{\epsilon}$ if ${\text{col}(\bm A) \cap \text{col}(\bm B) \neq \emptyset}$.

To perturb an equimolar coexistence point via Method II, we rotate $\vec{\mu}\in \mathcal{R}^N$ by a small angle $\theta$ in the plane specified by two orthonormal vectors $\hat n_1$ and $\hat n_2$.
To this end, we define the rotation
\begin{align}
  R_{\hat n_1 \hat n_2}(\theta) &= I + (\hat n_2 \hat n_1^\top - \hat n_1 \hat n_2^\top) \sin\theta \nonumber \\
  &\qquad + (\hat n_1 \hat n_1^\top + \hat n_2 \hat n_2^\top) (\cos\theta + 1)
\end{align}
such that the perturbed chemical-potential vector is
\begin{equation}
  \vec{\mu} = \vec{\mu_0} + \Delta \vec{\mu} = R_{\hat n_1 \hat n_2}(\theta) \vec{\mu}_0.
\end{equation}
For example, if we choose ${\hat n_1 = (1, 0, \ldots, 0)}$ and ${\hat n_2 = (0, 1, \ldots, 0)}$, then we only perturb $\mu_1$ and $\mu_2$, leaving the chemical potentials of the other components unchanged.
In practice, we apply a sequence of rotations with uniformly distributed random angles in the range $[0, 0.005\pi)$ for all pairs of axes, and then solve for the perturbed interaction matrix via $\bm A \Delta \vec{\epsilon} = \vec{b}$.
Finally, since this approach is only accurate to linear order, we fine-tune the coexistence point using  the nonlinear phase coexistence solver described in \appref{app:nonlinear-coex}.
Applying this method repeatedly produces a random walk in $\bm\epsilon$-space, in which every interaction matrix corresponds to a coexistence point with the target phase-diagram topology but, in general, non-equimolar condensed-phase volume fractions (\figref{appfig:perturbation}D).
As noted above, the changes in the condensed-phase volume fractions relative to the initial equimolar coexistence point tend to be correlated across phases and components when the target-phase-diagram manifold in $\bm\epsilon$-space is low dimensional.

\section{Free-energy calculations in a molecular model with pair potentials}
\label{app:fe-calc}

\subsection{Model definition}
\label{app:lattice-model}

We consider a multicomponent three-dimensional square lattice-gas model in which solute molecules interact via short-ranged pair potentials.
Specifically, if two lattice sites separated by a distance $r$ are occupied by solute molecules of types $i$ and $j$, then the additive contribution to the potential energy is
\begin{equation}
  u_{ij}(r) = \begin{cases}
    r < a & \infty \\
    a \le r < 2a & \left(\frac{10}{z}\right) \epsilon^{\text{MF}}_{ij} \\
    r \ge 2a & 0,
  \end{cases}
\end{equation}
where $a$ is the lattice constant, $z = 26$ is the number of neighboring lattice sites within a distance $1 \le r / a < 2$, and $\bm\epsilon^{\text{MF}}$ is the designed interaction matrix obtained from the regularized $L=1$ Flory--Huggins SDP.
All solute molecules of type $i$ are assigned a chemical potential $\mu_i$.
Vacant lattice sites, which represent solvent, are non-interacting and have chemical potential $\mu_0 = 0$.

\subsection{Free-energy calculations at phase coexistence}

We use grand-canonical Monte Carlo simulations~\cite{frenkel2001understanding} to calculate the grand-potential landscape (also referred to as the ``free-energy landscape'' in what follows) at coexistence in this lattice model.
Following the method described in~\refcite{jacobs2013predicting} and \refcite{jacobs2021self}, we define an order parameter $\Delta\phi_{0\alpha}$ between the dilute phase and the $\alpha$ condensed phase,
\begin{equation}
  \Delta\phi_{0\alpha}(\vec\phi) \equiv (\vec\phi - \vec\phi^{(0)}) \cdot \hat\nu_{0\alpha},
\end{equation}
where ${\hat\nu_{0\alpha} \equiv (\vec\phi^{(\alpha)} - \vec\phi^{(0)}) / |\vec\phi^{(\alpha)} - \vec\phi^{(0)}|}$ and $\vec\phi^{(0)}$ and $\vec\phi^{(\alpha)}$ are the volume fractions at the grand-potential minima in the dilute and $\alpha$-phase free-energy basins, respectively.
To sample trajectories that reversibly transit between these two free-energy basins, we add a constraining potential in directions of concentration space orthogonal to $\hat\nu_{0\alpha}$,
\begin{equation}
  U_{0\alpha}(\vec\phi) \equiv k_\perp \big|(\vec\phi - \vec\phi^{(0)}) - [(\vec\phi - \vec\phi^{(0)}) \cdot \hat\nu_{0\alpha}] \hat\nu_{0\alpha}\big|^6.
\end{equation}
The efficiency of the simulation is improved by proposing particle exchanges from a lattice site occupied by a particle (or a vacancy) of type $i$ to a particle (or a vacancy) of type $j$ with probability
\begin{equation}
  p_{\text{gen}}(i \rightarrow j) = \begin{cases}
    0.5 &\text{if }j\text{ is a vacancy} \\
    \frac{0.5 - 0.01}{M^\alpha} &\text{if }j\text{ is enriched in phase }\alpha \\
    \frac{0.01}{N - M^\alpha} &\text{if }j\text{ is depleted in phase }\alpha
  \end{cases}
\end{equation}
and then accounting for $p_{\text{gen}}$ in the Metropolis acceptance criteria~\cite{frenkel2001understanding}.

We first perform Wang--Landau simulations~\cite{wang2001efficient,jacobs2021self} to compute the projected free-energy landscape, $F_{0\alpha}(\Delta\phi_{0\alpha})$, under the combined potential, ${\mathcal{H}_{\text{LG}} + U_{0\alpha}}$,
\begin{widetext}
\begin{equation}
  F_{0\alpha}(\Delta\phi') = -\log \sum_x \delta\left\{\Delta\phi_{0\alpha}[\vec\phi(x)] - \Delta\phi'\right\} \exp\left\{-\mathcal{H}_{\text{LG}}(x) - U_{0\alpha}[\vec\phi(x)]\right\} + \text{const}.,
\end{equation}
\end{widetext}
where $x$ represents a lattice configuration and $\mathcal{H}_{\text{LG}}(x;\vec\mu,\bm\epsilon)$ is the multicomponent lattice-gas Hamiltonian for the model described in \appref{app:lattice-model}.
We use an $L \times L \times L$ periodic lattice with ${L = 6}$ and ${k_\perp = 40^6}$.
Next, we perform multicanonical simulations~\cite{berg1992multicanonical}, using $-F_{0\alpha}[\vec\phi(x)]$ as a biasing potential to ``flatten'' the free-energy barrier separating the dilute and $\alpha$ phases.
We then use MBAR~\cite{shirts2008statistically} to combine samples from the $K$ multicanonical simulations, one for each condensed phase.
Reweighting the combined samples to the unbiased distribution, in which the probability of a lattice configuration $x$ is proportional to $\exp[-\mathcal{H}_{\text{LG}}(x)]$, yields the grand potential landscape, $\Omega(\vec\phi; \vec\mu)$, from which we can calculate the grand-potential differences between pairs of free-energy basins.
Finally, we determine the coexistence point by tuning $\vec\mu$ and reweighting $\Omega(\vec\phi; \vec\mu)$ such that the grand-potential differences among all pairs of phases vanish to within statistical uncertainty.
This final step is accomplished using the algorithm described in \appref{app:nonlinear-coex}.
The landscapes shown in \figref{fig:5} are constructed by projecting the grand potential at coexistence onto a two-dimensional space defined by the first two principal components of the reweighted simulation samples.

%


\end{document}